\documentclass[10pt]{article}
\usepackage{amsmath}
\usepackage{cite} 
\usepackage{indentfirst}
\usepackage{latexsym}
\usepackage{pstricks}

\long\def\ca#1\cb{} 

\newcommand{\ad}{^\dagger }

\newcommand{\AND}{{\small AND}}

\newcommand{\becs}{\begin{cases}}
\newcommand{\bem}{\begin{matrix}}
\newcommand{\blp}{\bigl(}

\newcommand{\brp}{\bigr)}
\newcommand{\bsk}{\bigskip }

\newcommand{\dya}[1]{|#1\rangle\langle#1|}

\newcommand{\encs}{\end{cases}}
\newcommand{\enm}{\end{matrix}}

\newcommand{\hf}{{\textstyle\frac{1}{2} }}

\newcommand{\inp}[1]{\langle#1|#1\rangle }
\newcommand{\inpd}[2]{\langle#1|#2\rangle }
\newcommand{\ket}[1]{|#1\rangle }

\newcommand{\lra}{\leftrightarrow }

\newcommand{\msk}{\medskip }

\newcommand{\mte}[2]{\langle#1|#2|#1\rangle }
\newcommand{\mted}[3]{\langle#1|#2|#3\rangle }

\newcommand{\od}{\odot }
\newcommand{\OR}{{\small OR}}
\newcommand{\ot}{\otimes }

\newcommand{\ra}{\rightarrow }
\newcommand{\Ra}{\Rightarrow }

\newcommand{\st}{\sqrt{2}}

\newcommand{\Tr}{{\rm Tr}}

\newcommand{\vbB}{\boldsymbol{\mid}}

\newcommand{\AC}{{\mathcal A}}
\newcommand{\BC}{{\mathcal B}}

\newcommand{\EC}{{\mathcal E}}
\newcommand{\FC}{{\mathcal F}}

\newcommand{\HC}{{\mathcal H}}

\newcommand{\MC}{{\mathcal M}}

\newcommand{\PC}{{\mathcal P}}
\newcommand{\QC}{{\mathcal Q}}

\newcommand{\SC}{{\mathcal S}}

\newcommand{\al}{\alpha }

\newcommand{\gm}{\gamma }
\newcommand{\Gm}{\Gamma }
\newcommand{\dl}{\delta }

\newcommand{\om}{\omega }

\def\outl#1{\par{\medskip\noindent\hspace*{0.1cm}\bf
      \mathversion{bold}#1\mathversion{normal}\smallskip} }
 \def\xa{} \def\xb{}  

\def\outl#1{}\def\xa{} \def\xb{}  

\ca
 \def\outl#1{\par{\medskip\noindent\hspace*{0.1cm}\bf
      \mathversion{bold}#1\mathversion{normal}\smallskip} }
 \long\def\xa#1\xb{}
 
\cb

\begin{document}

\ca
\centerline{\Large The New Quantum Logic}
\xa
\msk
\cb

\title{The New Quantum Logic}

\author{Robert B. Griffiths
\thanks{Electronic mail: rgrif@cmu.edu}\\ 
Department of Physics,
Carnegie-Mellon University,\\
Pittsburgh, PA 15213, USA}
\date{Version of 27 June 2014}
\maketitle  
\ca
\centerline{Robert B. Griffiths}
\centerline{Physics Department}
\centerline{Carnegie-Mellon University}
\vspace{.2cm}
\cb

\ca
\centerline{Version of 11 November 2013}
\vspace{0.5 cm}
\cb


\xb

\xa
\begin{abstract}
  It is shown how all the major conceptual difficulties of standard (textbook)
  quantum mechanics, including the two measurement problems and the (supposed)
  nonlocality that conflicts with special relativity, are resolved in the
  consistent or decoherent histories interpretation of quantum mechanics by
  using a modified form of quantum logic to discuss quantum properties
  (subspaces of the quantum Hilbert space), and treating quantum time
  development as a stochastic process.  The histories approach in turn gives
  rise to some conceptual difficulties, in particular the correct choice of a
  framework (probabilistic sample space) or family of histories, and these are
  discussed. The central issue is that the principle of unicity, the idea that
  there is a unique single true description of the world, is incompatible with
  our current understanding of quantum mechanics.
\end{abstract}
\xb

\tableofcontents
\bsk
\xa
\xb
	\section{Introduction}
\label{sct1}
\xa

\xb
\outl{Students find standard (textbook) QM difficult; textbooks lack some
basic principles}
\xa

The conceptual difficulties of standard quantum mechanics, defined as what one
finds in standard textbooks, are encountered by students in their very first
course in the subject.  Part of the problem is unfamiliar mathematics, but even
when the mathematics has been (more or less) mastered a serious problem remains
in relating the mathematical formalism to some physical understanding of
quantum systems. And it is no wonder that students are having difficulty if
even textbook writers do not really understand the subject, and are sometimes
bold enough to admit it \cite{Lloe12}.  Are there some basic
principles which are missing from the textbooks, ideas which were they included
therein would clear up quantum paradoxes?

\xb
\outl{Principles missing from textbooks are in widely ignored CH approach}
\xa

\xb
\outl{CH resolves all major Qm conceptual difficulties including measurement
  problem}
\xa

\xb
\outl{CH has been criticized. Mermin: CH vs SR difficulties $\lra$ elephant vs
 gnat}
\xa

The thesis of this paper is that there are such basic principles, which have
been around in some form or another for nearly thirty years, and they deserve
to be more widely known.  Although ignored in much of the current literature,
the (consistent or decoherent) histories approach appears capable of resolving
\emph{every} major conceptual difficulty of quantum mechanics, not least the
infamous measurement problem. To be sure, the histories approach has not been
completely ignored; a small but distinguished group of critics---who may
possibly outnumber the advocates---have not hesitated to point out what they
consider serious flaws; see the references given in \cite{Grff13}.  One of the
more generous of these critics, N.\ David Mermin, expressed what has probably
troubled many others when he made the following comparison with special
relativity, see p.~281 of \cite{Schl11} or p.~16 of \cite{Mrmn13}:
\begin{quote}
[But] I am disconcerted by the reluctance of some consistent historians to
acknowledge the utterly radical nature of what they are proposing. The
relativity of time was a pretty big pill to swallow, but the relativity of
reality itself is to the relativity of time as an elephant is to a
gnat.
\end{quote}

\xb
\outl{CH is radical. Feynman thot QM much more difficult than SR.}
\xa

\xb
\outl{Revised reasoning more radical than SR. More like earth beginning to 
move}
\xa

What the consistent historians are proposing is indeed radical, which should
surprise no one familiar with Feynman's famous remark that ``no one understands
quantum mechanics'' (p.~129 of \cite{Fynm65}). It occurs in a context where
Feynman makes it abundantly clear that he considers quantum theory much more
difficult than special relativity, though he does not quantify this by means of
a zoological analogy.  The historians' proposal to revise some of the rules of
reasoning which before the arrival of quantum mechanics were thought to apply
universally, both in scientific reasoning and in everyday human affairs, is
obviously more difficult to accept than the move from pre-relativistic to
relativistic physics.  It is much more like the transition our intellectual
ancestors made when they abandoned the notion that the earth is motionless at
the center of the universe in favor of the radical proposal that it moves
around the sun as well as spinning around its axis.  The question physicists
should be asking is not whether the ideas in the consistent histories approach
are radical, but rather whether they are internally consistent, and whether
they genuinely resolve the serious conceptual issues which have beset quantum
theory ever since its development in the mid 1920s.

\xb
\outl{Birkhoff and von Neumann Qm logic not very successful. Physicists not
  smart enough?}
\xa

However radical it may seem, the idea that quantum theory requires a new mode
of reasoning is itself not at all new. In 1936, just four years after the
appearance of von Neumann's famous book, Birkhoff and von Neumann \cite{BrvN36}
published their proposal for a \emph{quantum logic} as a replacement in the
quantum domain for ordinary propositional logic.  Through the years there has
been a continuing, albeit modest, research effort attempting to develop quantum
logic in hopes that it would lead to a solution of the quantum conceptual
difficulties.  Despite some early enthusiasm, e.g. \cite{Ptnm75}, this program
has not made a great deal of progress; for some discussion of the current
situation see \cite{Mdln05,Bccg09}.  It may be that we physicists are simply
not smart enough to reason in this fashion, and the quantum mysteries will have
to rest until the day when superintelligent robots (with access to quantum
computers?) can make sense of the quantum world. But will they be able (or even
want to) explain it to us?

\xb
\outl{New logic less radical, more successful than Birkhoff and von
  Neumann }
\xa

\xb
\outl{Outline of paper. QM difficulties in Tbl. 1, Sec.~\ref{sct2}; CH in 
Sec.~\ref{sct3} resolves these; conceptual problems of CH in Tbl. 2, 
Sec.~\ref{sct4};
conclusion in Sec.~\ref{sct5}.}
\xa

What is here called the \emph{new} quantum logic has the same motivation and
shares important ideas with the proposal of Birkhoff and von Neumann.  It is in
some respects a less radical break with conventional reasoning than the older
quantum logic, and has turned out to be much more useful in terms of allowing
human beings, including college seniors and beginning graduate students, to
understand the quantum world in a consistent and coherent way.  The present
paper is devoted to explaining how this approach resolves the major conceptual
problems of standard quantum mechanics listed in Table~\ref{tbl1} and discussed
in some detail in Sec.~\ref{sct2}.  Following that, Sec.~\ref{sct3} summarizes
the histories approach and how it addresses these difficulties.  Next,
Table~\ref{tbl2} lists and Sec.~\ref{sct4} discusses various conceptual
problems raised by the histories approach itself.  A brief conclusion follows
in Sec.~\ref{sct5}.

\xb
\section{Quantum Conceptual Difficulties}
\label{sct2}
\xa

\xb
\outl{Table with Conceptual Difficulties of QM}
\xa

Table~\ref{tbl1} is a list of major conceptual difficulties of standard quantum
mechanics; that is, the treatment currently found in most textbooks.  These are
topics which have given rise to a lengthy and continuing discussion in the
quantum foundations literature. While no such list can claim to be complete,
the author believes that most of the significant interpretational problems fall
in one or another of these categories.

\xb
 \begin{table}[h]
 \caption{Major Conceptual Difficulties of Quantum Mechanics}
 \label{tbl1}
\begin{center}
 \begin{tabular}{l l l}
\hline\vspace{-1.1ex}\\
 1. & \multicolumn{2}{l}{ Meaning of the wave function}\\
    & a. & Ontological\\
    & b. & Time development\\
    & c. & Epistemological\\[1ex]\hline\vspace{-1.1ex}\\
 2. & \multicolumn{2}{l}{ Measurements}\\
    & a. & Outcomes (pointer states)\\
    & b. & What was measured?\\
    & c. & Wave function collapse\\[1ex] \hline\vspace{-1.1ex}\\
 3. &   \multicolumn{2}{l}{Inteference}\\
    & a. & Particle vs.\ wave\\
    & b. & Delayed choice\\[1ex]\hline\vspace{-1.1ex}\\
 4. & \multicolumn{2}{l}{Locality}\\
    & a. & Bell inequalities\\
    & b. & GHZ and Hardy\\[1ex] \hline
 \end{tabular}
\end{center}
 \end{table}
\xa

\xb
\subsection{Meaning of a wave function}
\label{sbct2.1}
\xa

\xb
\outl{Ontological: Wave function like a point in phase space, a quantum beable}
\xa

Students are taught that a wave function or wave packet or ket in the quantum
Hilbert space is analogous to a point in a classical phase space; e.g., for
a single particle it contains information about both the position and the
momentum.  Wave packets have a unitary time development governed by the
Hamiltonian through Schr\"odinger's equation,
and under suitable circumstances the wave packet can be seen to ``move''
somewhat like a a classical particle---one thinks of the well-known Ehrenfest
relations.  Thus it is rather natural to conclude that the wave function
represents whatever it is in the quantum world that is ``really there,'' a
\emph{beable} in Bell's terminology \cite{Bll04}.  Let us call this the
\emph{ontological} perspective.

\xb
\outl{Epistemic: Wave fn $\ra$ probabilities, which can suddenly change. It
  provides information, but information about what? How is discontinuous change
related to Schr Eq?}
\xa

There are other circumstances in which a wave function seems to play a
different role.  It can be used to calculate probabilities of the outcomes of a
measurement.  Students learn that the process of measurement makes a wave
function collapse.  This can, it seems, take place instantly, which for a wave
function with significant extension in space might violate special
relativity. It is certainly contrary to the unitary continuous time development
induced by Schr\"odinger's equation.  Probabilities can be instantly updated
according to new knowledge, and relativity theory need not be violated in such
updating. So if a wave function is just a means of calculating the probability
of something, there is no reason why it should not suddenly change. Such is the
\emph{epistemic} understanding of wave functions: rather than actually
representing the physical state of affairs they only provide \emph{information}
about a quantum system.  But what is this information \emph{about}?  Presumably
quantum theory is able in principle, even if in practice the calculations may
be very difficult, to tell us something about what is going on in systems which
have been probed experimentally leading to the conclusion that classical
mechanics is not an adequate representation of atomic systems.  Is there
something really \emph{there}, the way experimentalists seem to think, and if
so how is \emph{it} related to the wave function? And how is discontinuous
time development related to Schr\"odinger's equation?

\xb
\outl{Reconciling ontological \& epistemic a serious problem not helped by
  replacing $\ket{\psi}$ with $\rho$}
\xa

Reconciling the ontological and epistemic points of view is a serious
conceptual problem, and it does not disappear when one replaces a wave function
with a density operator, somewhat analogous to a classical probability
distribution.  A classical distribution provides a probability of something
definite, which either occurs or does not occur.  But what is the referent of a
quantum probability distribution?

\xb
\subsection{Measurements}
\label{sbct2.2}
\xa

\xb
\outl{Measurement problem: provide fully Qm description of entire process}
\xa

Measurements play a central role in textbook expositions of quantum theory.
The students are suspicious, and rightly so.  After all, the measuring
apparatus is itself constructed from a large collection of atoms whose behavior
is governed by quantum laws, and therefore it should be possible, at least in
principle, to describe the entire measuring process, both the system being
measured and the measuring apparatus, in fully quantum mechanical
terms. Providing an adequate and fully \emph{quantum} description constitutes
the infamous \emph{measurement problem} of quantum foundations.

\xb
\outl{Two measurement problems. \#1. Schr cat state for pointer. Decoherence 
does not help}
\xa

There are actually two distinct measurement problems. The \emph{first}, the one
most often discussed in the foundations literature, has to do with the fact
that a measuring process amplifies microscopic differences in such a way as to
make them macroscopic. In the dated but picturesque terminology of this field,
these difference are ultimately revealed through different positions occupied
by a macroscopic \emph{pointer} that indicates the measurement outcome. When
unitary time evolution is applied to both the microscopic system being measured
and the apparatus the result will often be a quantum wave function which is a
superposition of different macroscopic pointer positions. (See
Sec.~\ref{sbct3.5} below for a particular measurement model.)  How is such a
quantum state, nowadays often called a Schr\"odinger cat, to be understood?  Is
the pointer oscillating back and forth between different positions unable,
so-to-speak, to make up its mind?  Superficial invocations of decoherence do
not really resolve the problem \cite{Adlr03}.

\xb
\outl{Measurement problem \#2.  Infer prior property of measured system}
\xa

If one can somehow get the pointer to stop wiggling and settle down in a
definite position, the \emph{second} measurement problem remains: how is 
this position  related to the microscopic state of affairs that the apparatus
was designed to measure?  Experimenters typically claim that the outcomes of
their experiments tell them something about a prior state of affairs.  E.g., a
gamma ray was detected coming form a decaying nucleus, or a neutrino from a
distant supernova was detected by the apparatus.  This seems directly contrary
to the claim found in some textbooks that measurements tell one nothing about
what was there before the measurement took place.  

\xb
\outl{Proper analysis of measurements should explain wave fn collapse}
\xa

The third item under the measurement heading in Table~\ref{tbl1}, wave function
collapse, has already been mentioned in Sec.~\ref{sbct2.1}.  Obviously, an
adequate and fully quantum mechanical description of the measurement process
should provide some insight into why collapse can be a useful epistemic
perspective, or else replace it with something else which will accomplish
the same purpose.  

\xb
\subsection{Interference}
\label{sbct2.3}
\xa

\xb
\outl{Double slit, MZ interference.  Delayed choice, indirect measurement}
\xa

Double-slit interference leads to a well-known paradox in which one must
understand a quantum particle as a wave that is sufficiently delocalized that
it can in some sense pass through both of the slits in order to produce an
interference pattern. However, if detectors are placed immediately behind the
slits only one, not both, will be triggered, indicating that the particle
passed through only one slit.  And despite its wavelike character the particle
can arrive at a quite specific location in the interference region. Feynman's
superb discussion in Ch.~1 of \cite{FyLS65} can be recommended to any reader
who has not yet encountered it.  A very similar paradox occurs in a
Mach-Zehnder interferometer where a photon must in some sense be moving through
both arms in order to produce the expected interference at the second beam
splitter, whereas a measurement inside the interferometer will detect the
photon in just one arm, not both.  The paradox is even more striking in
Wheeler's delayed choice version \cite{Whlr78}, where the final beam splitter
in the Mach-Zehnder is either left in place or else suddenly removed at a time
when the photon has already passed through the first beam splitter.  In yet
another version \cite{ElVd93} the fact that one arm of the interferometer is
blocked can seemingly be detected by a photon which passes through the other
arm a long distance away.

\xb
\subsection{Locality}
\label{sbct2.4}
\xa

\xb
\outl{Bell inequalities, also Hardy \& GHZ, violate QM, which is supported by
  experiments. So something must be wrong with derivations of Bell
  inequalities,  etc.}
\xa

Bell inequalities \cite{Bll64b,Bll90c} apply to a situation where two quantum
particles, typically two photons, are prepared in an entangled state and
various measurements are used to determine the statistical correlations of some
of their properties.  By making certain assumptions about the presence of
physical properties in the particles before measurement, and assuming locality,
which is that influences only travel at a finite speed from one point to
another, Bell derived certain inequalities these correlations should satisfy.
The inequalities are violated by the predictions of quantum mechanics, and
numerous experiments of increasing precision all agree with quantum theory and
disagree with Bell's inequalities.  This has convinced most physicists that one
or the other of Bell's assumptions must be wrong.  Hardy's paradox
\cite{Hrdy92}, which also applies to correlations of properties of two
separated particles, is to some degree more straightforward than Bell's work as
it is easier to see that quantum predictions for the correlations, again in
accord with experiments, contradict what one might naively expect if
measurements reveal prior properties.  The paradox of Greenberger, Horne, and
Zeilinger \cite{GrHZ89,GHSZ90} is similar to Hardy's and preceded it in time,
but refers to three particles rather than two.  Once again, the predictions of
quantum mechanics are supported by experiment, suggesting that something must
be wrong with the reasoning that leads to these paradoxes.

\xb
\outl{Frequent claim: preceding implies there exist nonlocal Qm influences }
\xa

\xb
\outl{These influences cannot transmit info, so experimentally undetectable}
\xa

The claim has often been made that the only reasonable conclusion to be drawn
from the experimental violation of Bell inequalities and these other paradoxes
is that the quantum world must be nonlocal, and allow for instantaneous
interactions or influences between spatially separated systems, even in
situations where this conflicts with special relativity.  However, even those
who believe in the existence of such influences agree that they cannot be used
to transmit information, which conveniently makes them experimentally
unobservable.

\xb
\section{The New Quantum Logic}
\label{sct3}
\xa

\xb
\outl{Problems resolved using frameworks + stochastic time development }
\xa

\xb
\outl{Brief comments on contents of following subsections}
\xa

In resolving the conceptual difficulties listed in Table~\ref{tbl1} the
histories approach uses two main tools.  The first is the new quantum logic,
employed in Sec.~\ref{sbct3.1} to discuss physical properties of a quantum
system at a single instant of time, and extended to probabilities in
Sec.~\ref{sbct3.2}.  The second is stochastic time development, the
subject of Sec.~\ref{sbct3.3}.  Together these provide a way to
understand the dynamics of macroscopic systems using quasiclassical frameworks,
Sec.~\ref{sbct3.4}, and thus resolve both measurement problems, as
discussed in Sec.~\ref{sbct3.5}.  Interference and the locality of quantum
mechanics are taken up in Secs.~\ref{sbct3.6} and \ref{sbct3.7}. Some brief
remarks on approximations in Sec.~\ref{sbct3.8} complete the discussion of the
new logic and how it resolves quantum paradoxes.

\xb
\subsection{Properties}
\label{sbct3.1}
\xa

\xb
\outl{Old and new Qm logic: property $\lra$ subspace $\PC$, projector $P$}
\xa

\xb
\outl{Single $\ket{\psi}$ $\ra$ one-d subspace projector $[\psi]=\dya{\psi}$}
\xa

\xb
\outl{Subspaces of dimension greater than 1 also represent Qm properties}
\xa

The new quantum logic shares with its older counterpart a very fundamental
idea, consistent with but seldom sufficiently emphasized in quantum textbooks.
A quantum \emph{physical property}, something which can be true or false---such
as ``the energy is between $2$ and $3$ J''---is represented in quantum
mechanics by a (closed) \emph{subspace} $\PC$ of the quantum Hilbert space or,
equivalently, by the \emph{projector} $P$ (orthogonal projection operator) onto
this subspace.  Subspaces (or their projectors) represent the quantum ontology,
they are the mathematical counterparts of Bell's ``beables.'' Note that a wave
packet or any nonzero ket $\ket{\psi}$ in the Hilbert space corresponds to, or
generates, a one-dimensional subspace consisting of all its multiples: kets of
the form $c \ket{\psi}$, where $c$ is any complex number.  When $\ket{\psi}$ is
normalized we denote the projector onto this subspace by $[\psi] =
\dya{\psi}$. Used in this way a ket or wave function has an ontological
meaning.  However, kets can also play an epistemological role, as discussed
below in Secs.~\ref{sbct3.2} and \ref{sbct3.3}.  It is important to note that
subspaces of dimension greater than one also represent quantum properties.

\xb
\outl{Cl phase space $\Gm$, property $\PC$, indicator function $P(\gm)$
 $\lra$ Qm  projector }
\xa

A classical phase space $\Gm$ with a point in the phase space $\gm$
representing the actual physical state, provides a useful analogy for the
quantum Hilbert space.  A collection of points $\PC$ in $\Gm$ represents a
classical property, and this property is true for a given system if the point
$\gm$ representing its physical state is in the set $\PC$.  There is a
one-to-one correspondence between the subset $\PC$ and the corresponding
\emph{indicator function} $P(\gm)$, equal to 1 for $\gm\in \PC$ and 0
otherwise.  This classical indicator is thus analogous to a quantum projector,
whose eigenvalues are 1 and 0.

\xb
\outl{Cl physical variable = real valued fn on phase space}
\xa

\xb
\outl{Qm analog: observable $A=\sum_j a_j P_j$; $\{P_j\}$ form PDI }
\xa

\xb
\outl{(Property $A=a_j$) $\lra$ $P_j$. (Measurement of $A$) $\lra$ determine 
$P_j$}
\xa

In addition, classical mechanics employs various \emph{physical variables}
represented by real-valued functions on the phase space; e.g., energy,
momentum, angular momentum.  In quantum mechanics a physical variable, referred
to as an \emph{observable}, is represented by a Hermitian operator which can be
written in the form
\begin{equation}
A = \sum a_j P_j,\quad P_j =  P_j\ad = P_j^2,\quad \sum_j P_j = I,
\label{eqn1}
\end{equation}
where each eigenvalue $a_j$ of $A$ occurs but once in the sum: $j\neq k$
implies
$a_j\neq a_k$.  Here the $\{P_j\}$ are a collection of projectors that form a
\emph{projective decomposition of the identity operator} $I$, sometimes
called a projector-valued measure or PVM.  The property that the physical
variable $A$ takes on or possesses the value $a_j$, thus $A=a_j$, corresponds
to the projector $P_j$ or, equivalently, the subspace $\PC_j$ onto which $P_j$
projects.  While measurement will be discussed in more detail below in
Sec.~\ref{sbct3.5}, it is worth remarking that in quantum mechanics the
measurement of an observable is the same thing as measuring the corresponding
decomposition of the identity, determining which property $\PC_j$, equivalently
$P_j$, is, in fact true.

\xb
\outl{Negation, Cl, Qm. For Qm, use $\PC^\perp$ or $I-P$}
\xa

The negation of a classical property $\PC$ corresponds to the complementary
subset $\PC^c$ in the phase space: the set of points in $\Gm$ which are not in
$\PC$. Its indicator, $P^c$ or $\lnot P$, is $I-P$, where $I$ is the identity:
$I(\gm)=1$ for every $\gm\in\Gm$.  Von Neumann proposed that in quantum theory
the negation $\lnot P$ should be represented not by the set-theoretical
complement of the corresponding Hilbert subspace, but instead by its
\emph{orthogonal complement}, the collection $\PC^\perp$ of all kets which are
orthogonal to every ket in $\PC$. This is a subspace with projector $I-P$,
where $I$ is the identity operator on the Hilbert space.  This identification
is consistent with textbook quantum mechanics, though often it is not properly
discussed.

\xb
\outl{Expls: 1. Spin 1/2. Negation of $[z^+]$ is $[z^-]$, consistent with SG
  expt.  2. Harmonic oscillator}
\xa

Let us consider two examples.  For the two-dimensional Hilbert space
representing a spin-half particle the one-dimensional subspaces corresponding
to the orthogonal kets $\ket{z^+}$ and $\ket{z^-}$ (denoted by $\ket{0}$ and
$\ket{1}$ in quantum information theory), ``spin up'' and ``spin down'', are
associated with the projectors
\begin{equation}
 [z^+]=\dya{z^+},\quad [z^-]=\dya{z^-}.
\label{eqn2}
\end{equation}
As they sum to $I$ they are negations of each other, and together form a
projective decomposition of the identity.  If the spin is not ``down'' it is
``up'', consistent with what Stern and Gerlach observed in their famous
experiment. If $S_z=+1/2$ (in units of $\hbar$) is true, then $S_z=-1/2$ is
false, and vice versa. In the case of a quantum harmonic oscillator with energy
eigenstates $\ket{n}$, eigenvalues $(n+\hf)\hbar\om$, and projectors
$[n]=\dya{n}$, ``the energy is less than $2\hbar\om$'' is represented by the
projector $P= [0]+[1]$, and its negation, ``the energy is greater than
$2\hbar\om$'', by the projector $I-P = [3]+[4]+\cdots$. We shall return to
these examples later.

\xb
\outl{ $P\land Q$: old vs new Qm logic. New logic requires $PQ=QP$ to define 
$P\land Q$}
\xa

With reference to quantum properties and their negations the old and the new
quantum logic are identical.  The difference begins to emerge when one
considers the conjunction $P\land Q$, ``$P$ \AND\ $Q$'', of two
properties. Birkhoff and von Neumann defined it using the set-theoretical
intersection $\PC\cap\QC$ of the two subspaces, itself a (closed)
subspace. This has a precise analog in the classical phase space, where the
conjunction of two properties is represented by the intersection of the two
subsets $\PC$ and $\QC$.  However, if one follows the analogy of indicator
functions and projectors there is an important difference.  The indicator for a
classical conjunction $P\land Q$ is the product $PQ$ of the indicators.  But in
the quantum case the product $PQ$ of two projectors is a projector \emph{if and
  only if} $PQ=QP$, i.e., the two projectors \emph{commute}, in which case the
projectors, or the corresponding quantum properties, are said to be
\emph{compatible}.  For compatible properties the product $PQ$ of the
projectors projects on the subspace $\PC\cap\QC$.  However, when $P$ and $Q$ do
not commute, that is, the quantum properties or their projectors are
\emph{incompatible}, neither $PQ$ nor $QP$ is a projector, and there is no
simple relationship between either $PQ$ or $QP$ and the projector onto the
subspace $\PC\cap\QC$.  The new quantum logic differs from the old in that it
\emph{does not define} $P\land Q$ when the projectors do not commute; the
expression ``$P\land Q$'' in such a case is \emph{meaningless}: quantum
mechanics does not assign it a meaning.

\xb
\outl{Disjunction $P\lor Q$: old vs new Qm logic }
\xa

The old quantum logic defines the disjunction ``$P$ \OR\ $Q$'' (or both),
$P\lor Q$, as the \emph{span} of $\PC\cup\QC$, the union of the two collections
of kets. (Note that $\PC\cup\QC$ is in general not a subspace.)  The indicator
for the classical property $\PC\cup\QC$ is $P+Q-PQ$, and same expression works
for the quantum projector \emph{when $PQ=QP$}, but not otherwise.  Again, the
new quantum logic only defines the disjunction $P\lor Q$ when $P$ and $Q$
commute  Otherwise the disjunction is undefined, thus meaningless.

\xb
\outl{Syntactical rule prohibits combining incompatible properties; compare
$P\land\lor Q$ in std logic}
\xa

\xb
\outl{Negation of false statement is true, negation of meaningless statement is
meaningless}
\xa

The prohibition of conjunctions and disjunctions when $P$ and $Q$ do not
commute is an application of the \emph{single framework rule}, which plays a
central role in the new quantum logic, and about which more will be said in
Sec.~\ref{sbct3.2}.  This prohibition is a \emph{syntactical} rule governing
the combination of meaningful expressions (propositions or properties) to form
other meaningful expressions.  Thus, for example, in ordinary logic the
combination $P\land\lor\, Q$ has no meaning because it has not been constructed
according to the rules for meaningful expressions. In a similar way the new
logic forbids the combinations $P\land Q$ and $P\lor Q$ when $PQ\neq QP$; they
are meaningless. Note the difference between a statement which is meaningful
but \emph{false} and one which is \emph{meaningless}.  The negation of a false
statement is true, whereas the negation of a meaningless statement is equally
meaningless.

\xb
\outl{Spin half illustration: $[x^+]$ and $[z^+]$ don't commute.}
\xa

\xb
\outl{Old vs new logic for $[z^+]\land[x^+]$, $[z^-]\lor[x^-]$  }
\xa

\xb
\outl{Birkhoff and von Neumann had to remove distributive laws of logic}
\xa

The spin half example introduced above in \eqref{eqn2} provides a useful
illustration.  If we define
\begin{equation}
 \ket{x^+} = (1/\st)(\ket{z^+} + \ket{z^-},\quad
\ket{x^-} = (1/\st)(\ket{z^+} - \ket{z^-},
\label{eqn3}
\end{equation}
with $[x^+]$ and $[x^-]$ the projectors for the properties $S_x=\pm 1/2$, it is
easily checked that these do not commute with $[z^+]$ and $[z^-]$.  In the old
quantum logic the statement ``$S_z=+1/2$ \AND\ $S_x=+1/2$'' corresponds to the
zero operator, which is always false, and consistent with this its negation
``$S_z=-1/2$ \OR\ $S_x=-1/2$'' is always true. This last does not seem to make
much physical sense, and will indeed lead to a contradiction if one follows the
rules of ordinary reasoning---see Sec.~4.6 of \cite{Grff02c}.  Consequently, as Birkhoff
and von Neumann \cite{BrvN36} pointed out, it is necessary to modify the rules
of ordinary reasoning, by removing the distributive laws, in order to construct
a quantum logic free of contractions.
By contrast, in the new quantum logic, since ``$S_z=+1/2$ \AND\ $S_x=+1/2$'' is
meaningless, its negation is equally meaningless, and thus no contradiction
arises.  The approach in textbook quantum theory is to say that $S_x$ and $S_z$
cannot be simultaneously measured.  This is quite true, and one wishes that the
textbooks would go on and state the reason for this: even the most skilled
experimental physicists cannot measure that which does not exist!

\xb
\outl{New logic a subset of the old logic}
\xa

\xb
\outl{Advantage of new logic: ordinary rules of reasoning, probability}
\xa

The new quantum logic is in a sense a ``subset'' or restricted part of
the old quantum logic, as the former accepts only a special collection of the
formulas which are valid (constructed according to syntactical rules) for the
latter.  What is gained by adding this restriction is the ability to use the
ordinary rules of reasoning---and, as discussed below, the ordinary rules of
probability---to understand the quantum world without encountering
contradictions and paradoxes.

\xb
\subsection{Probabilities and the single framework rule}
\label{sbct3.2}
\xa

\xb
\outl{CH time development stochastic, not limited to measurements}
\xa

\xb
\outl{Sample space $\SC$, event algebra $\EC$, probability measure $\MC$;
finite or countable $\SC$ will do.}
\xa

In the histories interpretation the time development of a quantum system is a
stochastic process. Always, not just when measurements are being made.
Furthermore, the probabilities in question obey the standard rules of
probabilistic reasoning found in textbooks on probability theory (and which
ought to be found in quantum textbooks).  Three things are needed: a sample
space $\SC$ of mutually exclusive possibilities, one and only one of which is
true, a Boolean event algebra $\EC$, and a probability measure $\MC$
that assigns probabilities to the elements of $\EC$. For present
purposes we do not need sophisticated concepts.  A finite, or at most countable,
sample space will do very well, and $\EC$ can consist of all the subsets of
$\SC$, including $\SC$ and the empty set. 

\xb
\outl{Qm $\SC$ always a PDI. $\EC$ = sums of items in $\SC$, framework =
$\SC$ or $\EC$. All its projectors commute} 
\xa

\xb
\outl{Elements of $\SC$ are  \emph{elementary} events}
\xa

A quantum sample space is always a projective decomposition of the identity,
and any such projection can serve as a sample space. The event algebra consists
of all projectors which are sums of some of the projectors in the sample space,
including the zero projector 0 and the identity $I$.  Since the elements of a
projective decomposition of the identity commute with each other, so do all the
projectors in $\EC$. The term \emph{framework} will be used either for $\SC$ or
$\EC$; given the close relationship between the two this ambiguity should not
matter. When a distinction is important the elements of $\SC$ will be called
\emph{elementary} events or projectors.

\xb
\outl{(In)compatible frameworks defined using (non)commuting projectors}
\xa

\xb
\outl{Common refinement of compatible frameworks; incompatible frameworks
 do not have one}
\xa

Two frameworks with sample spaces $\{P_j\}$ and $\{Q_k\}$ are said to be
\emph{compatible} if every $P_j$ commutes with every $Q_k$; otherwise they are
\emph{incompatible}.  One arrives at exactly the same definition using
projectors belonging to the two event algebras: either they all commute
(compatible) or some do not (incompatible).  In the compatible case there is
always a smallest \emph{common refinement} of the two frameworks with a sample
space consisting of all nonzero products of the form $P_jQ_k$, with duplicates
eliminated.  The event algebra of the refinement includes all the
projectors in the two original event algebras.  Two incompatible frameworks do
not have a common refinement, and thus there is no way to combine the event
algebras.

\xb
\outl{Probability of event using sum of appropriate collection of $p_j$}
\xa

\xb
\outl{$p_j$ arbitrary except $p_j\geq 0,\,\sum p_j=1$. Qm time development
  gives certain conditional probs (Born rule). Probabilities can apply to
  a single system.}
\xa

To assign probabilities we start with a collection of nonnegative numbers
$\{p_j\}$, one for each projector in $\SC$, which sum to 1. Probabilities of
events in $\EC$ are calculated in the obvious way; e.g., $\Pr(P_2 + P_3 + P_5)
= p_2 + p_3 + p_5$.  Where do the $p_j$ come from?  For ordinary (classical)
probabilistic models they are simply parameters chosen by guesswork, or to
agree with experiment: there are no hard and fast rules.  The same is true in
quantum theory \emph{except} that for the time development of a closed system
the Born rule and its extensions, Sec.~\ref{sbct3.3}, provide certain
conditional probabilities which combined with appropriate assumptions (e.g., an
initial state) yield a probability distribution.  There is no reason quantum
probabilities should not be applied to single systems as is done for the
weather or to estimate the probability that the earth will collide with an
asteroid of a given size during the next millennium.

\xb \outl{SFR defined. It rules out $P$ \AND\ $Q$ if $PQ\neq QP$ (projectors
  in framework must commute)} \xa

The \emph{single framework rule} is a central principle of the new quantum
logic. What it says, in brief, is that a probabilistic calculation or a logical
argument must be carried out using a single framework, a single event algebra
generated by a single sample space, a single projective decomposition of the
identity.  Carrying out half of the reasoning or calculation using one framework
and then transferring the result to a different framework for additional
reasoning or calculations is prohibited. Since all projectors in a framework
commute with each other, this immediately rules out combinations of
noncommuting projectors using \AND\ or \OR\, as discussed in
Sec.~\ref{sbct3.1}.

\xb \outl{Harmonic oscillator example. $P=[0]+[1]$. Cannot use framework $\SC_1
  = \{P, I-P\}$ for $P\Ra$ energy is $1/2$ or $3/2\hbar\om$. Instead use
  $\SC_2=\{[0],[1],I-P\}$ } \xa

A useful illustration is provided by the harmonic oscillator,
Sec.~\ref{sbct3.1}, where $P=[0]+[1]$ is the property that the energy is not
greater than $2\hbar\om$.  It might seem obvious that if $P$ is true then the
energy is either $\hbar\om/2$ or $3\hbar\om/2$, the values when $n=0$ or 1.
However, this is \emph{not} a consequence of $P$ being true if we employ the
smallest framework $\FC_1$ which contains $P$, with sample space $\SC_1 = \{P,
I-P\}$.  The corresponding event algebra $\EC_1=\{0,P,I-P,I\}$ does not contain
either $[0]$ or $[1]$, so there is no way to discuss them. However, they are
included in the alternative framework $\FC_2$ with sample space
$\SC_2=\{[0],[1],I-P\}$, a refinement of $\FC_1$.  The event algebra $\EC_2$
includes both $[0]$ and $[1]$ as well as their sum $P$, and within this
framework one can use ordinary logic to infer that if $P$ is true then either
$[0]$ or $[1]$ is true, the energy is either $\hbar\om/2$ or $3\hbar\om/2$.

\xb
\outl{Difference $\FC_1$ vs $\FC_2$ is important; consider 
$\SC_3= \{[+],[-],I-P\}$}
\xa

\xb
\outl{Example of incorrect reasoning blocked by the SFR}
\xa

That the distinction between $\FC_2$ and $\FC_1$ is not just a matter of
nitpicking can be seen by introducing a third framework $\FC_3$ whose sample
space is $\SC_3 = \{[+],[-],I-P\}$, where $[+]$ and $[-]$ are projectors onto
the states $\ket{+}=(\ket{0}+\ket{1})/\st$ and $\ket{-}=(\ket{0}-\ket{1})/\st$.
Note that $[+]+[-]=P$, so $\FC_3$ is another refinement of $\FC_1$.  If we use
$\FC_3$ the truth of $P$ implies that the oscillator has either the property
$[+]$ or the property $[-]$, neither of which has a well-defined energy.  The
frameworks $\FC_2$ and $\FC_3$ are mutually incompatible because $[+]$ and
$[-]$ do not commute with $[0]$ and $[1]$---the relationship is formally the
same, if one restricts attention to the subspace $P$, as that between the $S_z$
and $S_x$ eigenstates of a spin-half particle.  The single framework rule
allows the use of either $\FC_2$ or $\FC_3$, but insists that they not be
combined.  Here is an argument that violates that rule: ``Let us suppose the
the oscillator is in the state $[+]$.  From this we infer (framework $\FC_3$)
that it possesses the property $P$.  But it is obvious (framework $\FC_2$) that
a system with the property $P$ has an energy of either $\hbar\om/2$ or
$3\hbar\om/2$.  Consequently the state $[+]$ has one of these two energies.''
It is this sort of reasoning, which in the quantum domain can lead to
paradoxes, that is blocked by the single framework rule.

\xb
\outl{Utility of $\FC_3$: oscillator in superposition of $\ket{0}$, $\ket{1}$}
\xa

But why would one ever want to use a framework such as $\FC_3$?  It is nowadays
possible to prepare a harmonic oscillator, either a mechanical oscillator or
the electromagnetic field inside a cavity, in a superposition of the ground and
first excited state, and $\FC_3$ might be useful in describing such a
situation.  Thus in quantum mechanics it is quite possible to say that ``the
energy is less than $2\hbar\om$'' \emph{without} implying that the energy is
equal to either of the two possible energies that are less than this value, and
this is precisely the significance of the projector $P$.

\xb
\outl{SFR means: Liberty, Equality, Incompatibility, Utility}
\xa

\xb
\outl{Physicist's choice of framework does not affect reality}
\xa

Four principles provide a compact summary of what the single framework rule
does and does not mean. First, the physicist has perfect Liberty to construct
different, perhaps incompatible, frameworks when analyzing and describing a
quantum system.  No law of nature singles out a particular quantum
framework as the ``correct'' one; from a fundamental point of view there is
perfect Equality among different possibilities.  However, the principle of
Incompatibility prohibits \emph{combining} incompatible frameworks into a
single description, or in employing them for a single logical argument leading
from premisses to conclusions.  The last principle is Utility: not every
framework is useful for understanding a particular physical situation or
addressing certain scientific questions.  In addition it is important to avoid
thinking that the physicist's choice of framework somehow influences reality.
Instead, quantum reality allows a variety of alternative descriptions, useful
for different purposes, which when they are incompatible cannot be
combined.

\xb
\subsection{Time development}
\label{sbct3.3}
\xa

\xb
\outl{Unitary vs stochastic time evolution}
\xa

\xb
\outl{Von Neumann, Everett, CH:stochastic evolution, textbooks}
\xa

Von Neumann's quantum mechanics had two distinct sorts of time evolution:
unitary evolution, based on Schr\"odinger's equation, and a separate stochastic
evolution associated with measurements, Sec.~V.1 of \cite{vNmn32b}. Few have
found this satisfactory, but devising something better has proven difficult. In
the Everett or many worlds interpretation \cite{Evrt57,DWGr73} there is only
unitary time development: a single unitarily evolving wave function of the
universe, or \emph{uniwave} in the terminology of \cite{Grff13}.  Proponents of
this approach then have to explain the probabilistic behavior of quantum
systems observed in the laboratory, a not altogether easy task. The histories
interpretation takes the opposite approach: all quantum time development is
\emph{stochastic}, and the deterministic Schr\"odinger equation is used to
calculate probabilities.  This is also what is done in textbooks, where physics
is extracted from the formalism using absolute squares of transition
amplitudes, though the whole matter is obscured through frequent (and
unnecessary) references to measurements.

\xb \outl{History $Y$: sequence of
  projectors (Qm properties) at successive times} \xa

\xb
\outl{$Y$ a projector on history Hilbert space $\breve\HC$}
\xa

\xb
\outl{Physical interpretation of $Y$: $F_0$ at $t_0$, $F_1$ at $t_1$\dots}
\xa

Stochastic time evolution requires a sample space with events at successive
times, and in the histories approach each event is a quantum property.  Thus
for a sequence of times $t_0<t_1<\cdots  t_f$ the sequence of properties
\begin{equation}
 Y = F_0\od F_1\od\cdots F_f,
\label{eqn4}
\end{equation}
where each $F_j$ is a projector, is a \emph{history} to which under appropriate
conditions one can assign a probability.  The $\od$ in \eqref{eqn4} indicates a
tensor product. (While it would be perfectly correct to use the standard symbol
symbol $\ot$, it is often helpful when considering the time development of a
quantum system possessing subsystems to employ a distinct symbol that separates
situations at different times.) The operator $Y$ is a projector on a subspace
of the \emph{history Hilbert space}
\begin{equation}
\breve\HC = \HC\od\HC\od\cdots \HC
\label{eqn5}
\end{equation}
formed from the tensor product of copies of the Hilbert space $\HC$ that
describes the system at a single time.  Its physical interpretation is that
that the (quantum) event $F_0$ occurred or, equivalently, the property $F_0$
was true at the time $t_0$, $F_1$ at the time $t_1$, and so forth.  

\xb
\outl{Sample space $\SC$ of elementary histories: projectors sum to $\breve I$}
\xa

\xb
\outl{Both $\SC$ \& $\EC$ are called ``family of histories''}
\xa

\xb
\outl{Prob of history in $\EC$: sum of probs of $\SC$ elements it contains}
\xa

\xb
\outl{One and only one elementary history actually occurs in any exptl run}
\xa

The sample space $\SC$ consists of a collection of orthogonal projectors of the
kind shown in \eqref{eqn4}, the \emph{elementary histories}, that sum to the
history identity,
\begin{equation}
\breve I = I\od I\od \cdots I,
\label{eqn6}
\end{equation}
and thus constitute a projective decomposition of $\breve I$.  The
corresponding event algebra $\EC$ consists of projectors which are sums of some
of the projectors that make up $\SC$, and the probability of any history in
$\EC$ is the sum of the probabilities of the elementary histories of which it
is composed.  The term ``family of histories'' is often employed in place
``framework'', and depending on the context can refer to either $\SC$ or $\EC$.
As in any probabilistic model, one and only one of the elementary histories,
which are mutually exclusive, occurs in any given situation or ``experimental
run.''

\xb
\outl{Family with fixed initial state $[\psi_0]$}
\xa

A simple but fairly useful family employs a set of elementary histories
\begin{equation}
 Y^\al = [\psi_0]\od P_1^{\al_1}\od P_2^{\al_2}\od \cdots  P_f^{\al_f},
\label{eqn7}
\end{equation}
where $[\psi_0]=\dya{\psi_0}$ is a fixed initial state at $t_0$, and at the
later time $t_m$ the projector $P_m^{\al_m}$, where $\al_m$ is a label not an
exponent, belongs to a fixed decomposition of the (single time) identity,
\begin{equation}
 \sum_{\al_m} P_m^{\al_m} = I,
\label{eqn8}
\end{equation}
and $\al = (\al_1,\al_2,\dots \al_f)$ is the label for the history $Y^\al$.  If
one includes along with the $Y^\al$ in \eqref{eqn7} a special history $Y^0 =
(I-[\psi_0])\od I\od I \od \cdots I$, the result is a family for which the
projectors add to $\breve I$, as required for a sample space.

\xb
\outl{Assigning probabilities for 2-time histories: Born rule using $T(t,t')$}
\xa

The rules for assigning probabilities to elementary histories of a closed
quantum system whose unitary time development is generated by a Hermitian
Hamiltonian are best explained using examples. The simplest situation is that
in which $f=1$, so only two times $t_0$ and $t_1$ are involved.  Let $T(t,t')$
be the unitary time development operator for the time interval from $t'$ to
$t$; for a time-independent Hamiltonian $H$ it is
$T(t_1,t_0)=\exp[-i(t_1-t_0)H/\hbar]$. At time $t_1$ assume that the projective
decomposition of the identity
corresponds to an orthonormal basis $\{\ket{\phi_1^k}\}$, and let $P_1^k =
  [\phi_1^k]$.  The Born rule assigns a conditional probability
\begin{equation}
  p_k = \Pr(P_1^k\vbB [\psi_0]) = |\mted{\phi_1^k}{T(t_1,t_0)}{\psi_0}|^2
\label{eqn9}
\end{equation}
to $P_1^k$ at $t_1$ given the initial state $[\psi_0]$ at $t_0$.
If we assign a probability of 1 to $[\psi_0]$ and 0 to $I-[\psi_0]$, which is
to say we assume the system starts at $t_0$ in the initial state $[\psi_0]$,
then $p_k$ is the probability assigned to the elementary history
$Y^k=[\psi_0]\od[\phi^k]$.

\xb
\outl{Textbooks: use uniwave $\ket{\psi(t)}$ to calculate Born probabilities}
\xa

Textbooks use the same rule, but tend to word it differently. Solving
Schr\"odinger's equation with initial state $\ket{\psi_0}$ at $t_0$ yields
\begin{equation}
\ket{\psi(t)} = T(t,t_0)\ket{\psi_0},
\label{eqn10}
\end{equation}
at time $t$. Setting $t=t_1$ allows us to write  $p_k$ in \eqref{eqn9} as
\begin{equation}
 p_k = |\inpd{\phi_1^k}{\psi(t_1)}|^2.
\label{eqn11}
\end{equation}
That is, students are taught to first calculate the uniwave \eqref{eqn10}, and
then use \eqref{eqn11} to find probabilities.  
(The other difference is that the textbooks generally refer to $p_k$ as the
probability of a (macroscopic) measurement outcome, the pointer position, rather
than as the probability of the (microscopic) property the measurement apparatus
was designed to measure.  The connection between the two will be discussed in
Sec.~\ref{sbct3.5} below.)

\xb
\outl{Two reasons why $\ket{\psi(t)}$ is not a physical property}
\xa

While this is an excellent calculational technique and
gives the right answers, it has unfortunately given rise to the idea
that the uniwave constitutes the fundamental quantum ontology, it represents
quantum reality.  There are two ways to see that this is mistaken.  First,
if one regards $[\psi(t_1)]$ as a quantum property, then it will not commute
with any $[\phi_1^k]$ for which $0<p_k<1$. Thus except in special cases the
single framework rule prevents 
$[\psi(t_1)]$ from being added to the set of
properties $\{[\phi_1^k]\}$ under consideration.  Second, note that
$p_k$ can be calculated using the formula
\begin{equation}
 p_k = |\inpd{\psi_0}{\hat\phi^k(t_0)}|^2;\quad
\ket{\hat\phi^k(t)} = T(t,t_1)\ket{\phi_1^k}.
\label{eqn12}
\end{equation}
That is, start with $\ket{\phi_1^k}$ at time $t_1$ and integrate
Schr\"odinger's equation backwards in time to obtain $\ket{\hat\phi^k(t)}$ at
$t=t_0$.  In this procedure (which is not particularly efficient, since to
obtain probabilities for several different $k$ requires integrating
Schr\"odinger's equation a comparable number of times) the uniwave never
appears, which shows that it was merely a convenient calculational tool.

\xb
\outl{$\ket{\psi(t)}$ is a pre-probability; its use is epistemic}
\xa

Referring to $\ket{\psi(t)}$ as a ``pre-probability'' (something used to
calculate a probability), as in Sec.~9.4 of \cite{Grff02c}, serves to emphasize
that, at least in the situation under consideration, it is \emph{not} to be
regarded as a quantum property, a genuine ``beable.''  (Similarly,
$\ket{\hat\phi^k(t)}$ in \eqref{eqn12} is a pre-probability.)
Consequently, the appropriate use of ``the wave function'' obtained by solving
Schr\"odinger's equation is, at least in general, epistemic: it is employed to
compute probabilities.  The claim, as found for example in
\cite{PsBR12,ClRn12,Hrdy13,PtPM13}, that the wave function cannot be used in
this way seems to be based on the use of classical hidden variables, which are
inconsistent with Hilbert-space quantum mechanics \cite{Grff13d}.

 \xb
\outl{Chain kets, consistency for histories of 3 or more times}
\xa

\xb
\outl{Reduces to previous formula for 2 times ($f=1$)}
\xa

When histories involve three or more times an extension of the simple Born rule
is needed in order to generate a consistent set of probabilities for quantum
histories.  The essential features appear already in the case of three times,
$f=2$, but it will be convenient to consider the general case of a family of
the type \eqref{eqn7}, For each elementary history $Y^\al$ define the
\emph{chain ket}
\begin{equation}
\ket{\al} = \ket{(\al_1,\al_2,\ldots \al_f)} 
 =P_f^{\al_f}T(t_f,t_{f-1})P_{f-1}^{\al_{f-1}}T(t_{f-1},t_{f-2})\cdots
P_1^{\al_1} T(t_1,t_0)\ket{\psi_0}.
\label{eqn13}
\end{equation}
Provided these chain kets are mutually orthogonal, which is to say
\begin{equation}
 \inpd{\al}{\al'} = 0 \text{ whenever } \al\neq \al' 
\label{eqn14}
\end{equation}
where $\al = \al'$ if and only if $\al_j=\al'_j$ for every $j$, then the
history $Y^\al$ is assigned the probability
\begin{equation}
\Pr(\al\vbB [\psi_0]) = \inp{\al}
\label{eqn15}
\end{equation}
conditioned on the initial state $[\psi_0]$.  The \emph{consistency conditions}
\eqref{eqn14} are automatically satisfied for histories only involving two
times, the case $f=1$, and then \eqref{eqn15} gives the same Born probability as
\eqref{eqn9}. However, for $f=2$ or more the restriction \eqref{eqn14}, which
depends both on the projectors making up the family and the unitary dynamics,
is needed and is not trivial.

\xb
\outl{Extended SFR. Families can be combined only if commensurate: consistency
 still  satisfied}
\xa

As noted previously, the single framework rule prohibits combining two sample
spaces when the projectors do not commute.  This applies also to combining two
families of histories: the history projectors must commute with each other for
the combination to be possible, and if this is not so we say the history
families are \emph{incompatible}. However, even if all the projectors in the
two families commute, it may be the case that the common refinement fails to
satisfy the consistency conditions, so there is no way to assign probabilities
to this family using the extended Born rule for a closed quantum system. In
that case we say the two families are \emph{incommensurate}.  It is then a
natural extension of the single framework rule to prohibit incommensurate as
well as incompatible families of histories. Or, to put it another way, the
usual rules of probabilistic quantum dynamics can only be applied to a single
consistent family, and results from two incommensurate, as well as from two
incompatible, families cannot be combined.  For an example of incommensurate
families see the discussion of families $\AC$ and $\BC$ for the Aharonov and
Vaidman three-box paradox given in Sec.~22.5 of \cite{Grff02c}.

\xb
\subsection{Quasiclassical frameworks}
\label{sbct3.4}
\xa

\xb
\outl{GMH proposal: Coarse-grained projectors, almost deterministic dynamics}
\xa

\xb
\outl{Exception: In chaotic Cl regime quasiclassical Qm histories will not be 
deterministic}
\xa

It has been argued by Omn\`es \cite{Omns99,Omns99b}, and Gell-Mann and Hartle
\cite{GMHr93,GMHr07,Hrtl11} (also see Ch.~26 of \cite{Grff02c}) that classical
mechanics for macroscopic systems emerges as a good approximation to a more
exact but unwieldy quantum description.  The idea is to use a
\emph{quasiclassical} quantum framework employing coarse-grained projectors
that project onto Hilbert subspaces of enormous, albeit finite, dimension,
suitably chosen so as to be counterparts of classical properties such as those
used in macroscopic hydrodynamics.  The stochastic quantum dynamics associated
with a family of histories constructed using these coarse-grained
quasiclassical projectors gives rise, in suitable circumstances, to individual
histories which occur with high probability and are quantum counterparts of the
trajectories in phase space predicted by classical Hamiltonian mechanics. There
are exceptions. For example, in a system whose classical dynamics is chaotic
with sensitive dependence upon initial conditions one does not expect the
quantum histories to be close to deterministic.

\xb 
\outl{Qcl framework not unique, but this not a great concern}
\xa

\xb
\outl{One qcl framework suffices for CP; SFR not needed, ordinary logic OK}
\xa

A quasiclassical family can hardly be unique given the enormous size of the
corresponding Hilbert subspaces, but this is of no great concern provided
classical mechanics is reproduced to a good approximation, in the sense just
discussed, by any of them.  Therefore all discussions which involve nothing but
classical physics can, from the quantum perspective, be carried out using a
single quasiclassical framework. As long as reasoning and descriptions are
restricted to this one framework there is no need for the single framework
rule, which explains why a central principle of quantum mechanics is absent
from classical physics.  And why ordinary propositional logic is
adequate for the macroscopic world of everyday affairs.

\xb
\subsection{Measurements}
\label{sbct3.5}
\xa

\xb
\outl{Measurement model of vN}

\xb
\outl{Particle, apparatus states; unitary 
evolution $\ket{\Psi_0}\ra\ket{\Psi_1}\ra\ket{\Psi_2}$}
\xa

A simple measurement model based on the one proposed by von Neumann in Ch.~V of
\cite{vNmn32b} will illustrate how the histories approach addresses the
measurement problems listed in Table~\ref{tbl1}. Suppose properties of a system
with Hilbert space $\HC_s$, hereafter referred to as a ``particle'', are to be
measured by an apparatus, Hilbert space $\HC_m$, with $\HC_s\ot\HC_m$ the
Hilbert space of the combined closed system.  Let
\begin{equation}
 \ket{s_0} = \sum_j c_j \ket{s^j},
\label{eqn16}
\end{equation}
be the initial state of the particle, where $\{\ket{s^j}\}$ is an orthonormal
basis of $\HC_s$, and $\ket{m_0}$ the ``ready'' state of the apparatus at the
initial time $t_0$. During the interval from $t_0$ to $t_1$ the particle and
apparatus do not interact, so we set $T(t_1,t_0) = I = I_s\ot I_m$
corresponding to a trivial dynamics. For the time interval from $t_1$ to $t_2$
during which they interact we assume that
\begin{equation}
T(t_2,t_1) \blp \ket{s^j}\ot\ket{m_0}\brp = \ket{s^j}\ot \ket{m^j},
\label{eqn17}
\end{equation}
where the $\ket{m^j}$, associated with different pointer positions, are
normalized and mutually orthogonal.  Thus under unitary time evolution the
initial state
\begin{equation}
\ket{\Psi_0} = \ket{s_0}\ot\ket{m_0}
\label{eqn18}
\end{equation}
develops into
\begin{equation}
 \ket{\Psi_1} = T(t_1,t_0)\ket{\Psi_0} = \ket{\Psi_0},\quad
 \ket{\Psi_2} = T(t_2,t_1)\ket{\Psi_1} = \sum_j c_j\ket{s^j}\ot\ket{m^j}
\label{eqn19}
\end{equation}
at the times $t_1$ and $t_2$. 

\xb
\outl{Unitary family $\FC_u$ cannot be used to discuss measurement 
outcomes}
\xa

\xb
\outl{First measurement problem insoluble in interpretations using uniwave}
\xa

Now consider various history families of the form \eqref{eqn7}, with the
initial state $[\Psi_0]=\dya{\Psi_0}$ at time $t_0$ given by \eqref{eqn18}.
One possibility is unitary time development:
\begin{equation}
 \FC_u:\;\;[\Psi_0]\;\od\; \{[\Psi_1],I-[\Psi_1]\}\;\od\; 
\{[\Psi_2], I-[\Psi_2]\}
\label{eqn20}
\end{equation}
for times $t_0<t_1<t_2$, with different histories in the sample space
constructed by choosing one of the projectors inside the curly brackets at each
of the later times.  Since the events $I-[\Psi_1]$ and $I-[\Psi_2]$ occur with
zero probability they can be ignored, and the single history $[\Psi_0]\od
[\Psi_1]\od [\Psi_2]$ occurs with probability 1. While $\FC_u$ is perfectly
acceptable as a family of quantum histories, it cannot be used to discuss
possible outcomes of the measurement because it does not include the projectors
$\{[m^j]\}$ for the pointer positions at time $t_2$, nor can it be refined to
include them, because $[\Psi_2]$ will not commute with some of the $[m^j]$,
assuming at least two of the $c_j$ in \eqref{eqn16} are nonzero. Thus the first
measurement problem cannot be solved if all time development is unitary.  This
is a basic difficulty facing all quantum interpretations that make the uniwave
fundamental to their ontology.

\xb
\outl{Family $\FC_1$ with pointer basis at $t_2$ resolves first measurement
  problem; $\ket{\Psi_2}$ can serve as pre-probability}
\xa

The histories approach can solve the first measurement problem by
replacing $\FC_u$ with the family
\begin{equation}
 \FC_1:\;\;[\Psi_0]\;\od\;[\Psi_1] \;\od\; \{[m^j]\}.
\label{eqn21}
\end{equation}
That is, the different histories agree at $t_0$ and $t_1$, but correspond to
different pointer positions at $t_2$.  The alternative $I-[\Psi_1]$, which
occurs with zero probability, has been omitted at $t_1$, and at time $t_2$ we
employ the usual physicist's convention that $[m^j]=\dya{m^j}$ means
$I_s\ot[m^j]$ on the full Hilbert space $\HC_s\ot\HC_m$. An additional
projector $R'=I -\sum_j [m^j]$ should be included at the final time in
\eqref{eqn12} so that the total sum is the $I$, but, again, it has probability
zero. While $[\Psi_2]$ cannot be one of the properties at time $t_2$ in family
$\FC_1$, see the discussion of $\FC_u$ above, it can be used as a
pre-probability (see the discussion following \eqref{eqn12}) to calculate
probabilities of the different pointer positions at time $t_2$:
\begin{equation}
\Pr([m^j]_2) = \Tr(\;[\Psi_2]\;[m^j]\;) = \mte{\Psi_2}{\,[m^j]\,}.
\label{eqn22}
\end{equation}
(Here and below we omit from $\Pr()$ the condition $[\Psi_0]$ at $t_0$, as it
applies in all cases.)

\xb
\outl{Family $\FC_2$ solves 2d measurement problem: infer prior state from
  measurement outcome}
\xa

\xb
\outl{Textbook `` probability of measurement outcome'' correct but 
confusing}
\xa

In order to relate the measurement outcome, the pointer position, to a prior
property of the measured particle and thus solve the second measurement problem
yet another family is needed:
\begin{equation}
 \FC_2:\;\;[\Psi_0]\;\od\; \{ [s^j]\}\;\od\; \{[m^k]\}.
\label{eqn23}
\end{equation}
The decomposition $\{ [s^j]\}$ at $t_1$ refers to
properties of the particle, $[s^j]$ means $[s^j]\ot I_m$, 
without reference to the apparatus.  It is straightforward to show that
 $\FC_2$ is consistent, leading to a joint probability
distribution
\begin{equation}
\Pr(\,[s^j]_1,[m^k]_2) = |c_j|^2 \dl_{jk},
\label{eqn24}
\end{equation}
where the subscripts on $[s^j]$ and $[m^k]$ indicate the time.  The marginals
are:
\begin{equation}
 \Pr([s^j]_1) = \Pr([m^j]_2) = |c_j|^2.
\label{eqn25}
\end{equation}
Thus if $|c_k|^2 > 0$ the conditional probability
\begin{equation}
 \Pr([s^j]_1 \vbB [m^k]_2) = \dl_{jk}
\label{eqn26}
\end{equation}
implies that from the (macroscopic) measurement outcome or pointer position
$[m^k]$ at time $t_2$ we can infer, using standard statistical inference, that
the particle had the (microscopic) property $[s^k]$ at the earlier time $t_1$.
This solves the second measurement problem.  And because the probability of
the $[s^j]$ at $t_1$ is the same as $[m^j]$ at $t_2$, textbooks in which
students are taught to calculate $|c_j|^2$ for the particle alone and then
ascribe the resulting probability to the outcome of a measurement, are not
wrong.  They would be less confusing if they provided a proper quantum
analysis of the measurement process, such as given here.

\xb
\outl{CH allows retrodiction so sometimes confused with HV approaches}
\xa

Because it solves the second measurement problem, a measurement actually
\emph{measures} something, the histories approach is sometimes confused with
the sorts of \emph{hidden variables} approach studied by Bell and his
followers.  However, the basic hypothesis underlying most hidden variables
schemes is that \emph{every} property which could possibly be measured already
``exists'' in some sense in the particle before measurement.  This leads to
various difficulties such as the Bell-Kochen-Specker paradox, for which see the
discussion in Sec.~22.1 of \cite{Grff02c}.  Suppose, for example, that a
measurement is to be carried out on a spin half particle. Since this might be a
measurement of $S_x$ or of $S_y$ or of $S_z$, it then seems natural to suppose
that all three values are somehow ``present'' in the particle before it is
measured. But this is contrary to Hilbert space quantum mechanics, since the
different projectors do not commute; see the discussion in Sec.~\ref{sbct3.1}.
The histories approach avoids the Bell-Kochen-Specker paradox by applying the
single framework rule \cite{Grff00b}.  This point will come up again in the
discussion of locality in Sec.~\ref{sbct3.7} below.  For the same reason the
histories approach rejects the notion that quantum mechanics is contextual; see
the detailed discussion in \cite{Grff13b}.

\xb
\outl{$\FC_u$,$\FC_1$, $\FC_2$ all legitimate, but have different utility}
\xa

It is to be noted that all three history families or frameworks, $\FC_u$,
$\FC_1$, and $\FC_2$ employed above satisfy the consistency conditions and thus
provide legitimate quantum descriptions.  There is no reason a priori to prefer
one to another; at a fundamental level there is Equality.  However, Utility
plays a significant role. If measurement outcomes (pointer positions) at $t_2$
are under discussion, $\FC_u$ is unsatisfactory, as its event algebra does not
contain them.  Both $\FC_1$, and $\FC_2$ allow discussion of measurement
outcomes, but if one is interested in how the outcomes are related to the
properties the device was designed to measure, the $[s^j]$ at time $t_1$,
$\FC_2$ makes this possible, whereas $\FC_1$ does not.

\xb \outl{$\FC_1$, $\FC_2$ both useful if Alice prepares $S_x$ \& Bob measures
  $S_z$} \xa

This does not mean that $\FC_2$ is always the ``right'' framework. Consider a
situation in which Alice prepares a spin half particle with $S_x=+1/2$ and
sends it through a field-free region to an apparatus that Bob has set up to
measure $S_z$, and which yields the value $S_z=-1/2$. At the intermediate time
the value of $S_x$ may be of interest to Alice---did the preparation device
work as intended?---in which case the $\FC_1$ family is appropriate.  However,
if Bob is concerned with how his apparatus has functioned, $\FC_2$ is more
relevant.  Both frameworks are valid tools for quantum analysis, and they could
both be used by the same person, e.g., someone who sets up both the preparation
and the measurement apparatus.  The restriction which the new logic imposes is
that they \emph{cannot be combined} into a single description

\xb
\outl{Family $\FC_3$: later particle state correlated with pointer position 
$\lra$  wave function collapse}
\xa

\xb
\outl{Collapse not an independent principle. Family $\FC_3$ is a
  \emph{preparation} }
\xa

  It is worth mentioning yet another family
\begin{equation}
 \FC_3:\;\;[\Psi_0]\;\od\;[\Psi_1] \;\od\; \{[s^k]\ot[m^j]\},
\label{eqn27}
\end{equation}
which is similar $\FC_1$ except that at time $t_2$ we have added the final
particle states to the description. It is easy to show, using $\ket{\Psi_2}$
from \eqref{eqn19} as a pre-probability, that the probability of any history
with $k\neq j$ is 0, and hence if $c_j$ in \eqref{eqn16} is nonzero,
\begin{equation}
 \Pr([s^k]_2\vbB [m^j]) = \dl_{jk}
\label{eqn28}
\end{equation}
That is to say, if at time $t_2$ the pointer is in the position $[m^j]$ the
particle is in the state $[s^j]$.  This is von Neumann's (and the textbooks')
``wave function collapse''.  But now it is simply an ordinary probabilistic
inference using a conditional probability, so there is nothing at all
mysterious about it, and it definitely does not have to be added to quantum
theory as an independent principle.  Other cases of wave function collapse
(when it is being properly used) can also be replaced by conditional
probabilities, thus eliminating another of the conceptual difficulties in
Table~\ref{tbl1}.  Note that $\FC_3$ is a family useful for analyzing a
\emph{preparation} procedure for producing a particle in a well-defined initial
state.

\xb
\outl{More realistic measurement models discussed in Ch.~17 of CQT}
\xa

Our discussion has employed various simplifications not present in real
measurements.  In particular, pointer positions will always be associated with
macroscopic properties, thus projectors onto enormous subspaces and not the
pure states $\ket{m^k}$ assumed above. Similarly, the initial state of a
macroscopic apparatus should be described using a macroscopic projector, or an
appropriate density operator.  The discussion given above is extended in Ch.~17
of \cite{Grff02c} to include these more realistic features, along with
irreversible (in the thermodynamic sense) behavior of the apparatus.  None of
the conclusions discussed above is undermined by this extension.

\xb
\subsection{Interference}
\label{sbct3.6}
\xa

\xb
\outl{Double slit, Mach-Zehnder paradoxes; latter discussed in detail in CQT
  Ch.~13 }
\xa

\xb
\outl{End results of CQT analysis applied to double slit}
\xa

The double slit and the similar Mach-Zehnder paradoxes were introduced in
Sec.~\ref{sbct2.3}.  The histories approach in the case of the Mach-Zehnder
interferometer is discussed in considerable detail in Ch.~13 of \cite{Grff02c},
and the same principles apply to the double slit.  Here are the end results of
that analysis.  A family of histories referring to the particle needs to
include both its existence in a coherent state $\ket{\psi_0}$ at a time $t_0$
before it encounters the slit system, and then its presence in a reasonably
compact region in the interference zone at a later time $t_2$.  One can then
argue that a family of histories in which the particle passes through one or
the other of the two slits at the intermediate time $t_1$ fails to satisfy the
consistency conditions, and thus cannot be discussed in appropriate
(probabilistic) quantum terms, in agreement with Feynman's conclusion based on
(excellent) physical intuition.  There is, on the other hand, an alternative
family in which the particle does pass through a definite slit and nothing is
said about what happens later.  Still another possibility is that the particle
passes through a definite slit and the quantum detectors in the later
interference region are left in a macroscopic quantum superposition
(Schr\"odinger cat) state.  Various possibilities are discussed in Ch.~13 of
\cite{Grff02c} using simplified models which are very useful for gaining a
better intuitive understanding of quantum interference.

\xb 
\outl{Consistency conditions interpreted as ``absence of interference''
  does not prevent CH discussion of situations where Qm interference occurs.}
\xa

The consistency condition \eqref{eqn14}, the requirement that the inner product
of chain kets for different elementary histories be zero, can be understood as
requiring an ``absence of interference'' when calculating probabilities.  This
should not be misinterpreted to mean that the histories approach cannot be
applied to physical situations, such as the double slit, where quantum
interference is central to the phenomena under consideration. Instead, for any
given physical situation, whether or not there is some form of interference,
the histories approach and in particular the consistency condition singles out
physically sensible and consistent ways of discussing what is going.

\xb
\subsection{Locality}
\label{sbct3.7}
\xa

\xb
\outl{Analogy of colored slips of paper sent by Charlie to Alice, Bob}
\xa

The following analogy shows how the histories approach counters the widespread
claim that quantum mechanics is nonlocal because it violates Bell inequalities.
Charlie in Chicago places a red slip of paper in an opaque envelope, a green
slip in another, and shuffles the two before mailing one to Alice in Atlanta
and the other to Bob in Boston.  Knowing the protocol followed by Charlie, if
Alice opens her envelope and sees a red slip of paper she can immediately
conclude that Bob's envelope contains (or contained) a green slip.  This
inference has nothing to do with whether Bob opens the envelope earlier or
later than Alice, or simply throws it away unopened.  Furthermore, Alice's
``measurement'' by opening and looking in the envelope has absolutely no
influence on the color, or any other property, of the slip in Bob's
envelope.

\xb
\outl{Spin singlet, Alice measures $S_{az}$, can infer $S_{bz}$}
\xa

How are things different if Charlie prepares two spin half particles $a$ and
$b$ in a singlet state, and pushes a button that sends $a$ towards Alice's
apparatus set up to measure $S_{az}$ for this particle, and $b$ towards Bob's
apparatus set up to measure $S_{bz}$?  If Alice's measurement outcome,
indicated by a suitable pointer, corresponds to $S_{az}=+1/2$ she is entitled,
as a competent experimentalist who knows how her equipment functions, to infer
that particle $a$ had this property before the measurement. By using a suitable
framework that includes both $S_{az}$ and $S_{bz}$ values at this earlier time,
and knowing the protocol followed by Charlie, she can infer the value 
$-1/2$ for $S_{bz}$.

\xb \outl{Framework: $\ket{\Psi_0}$ at $t_0$ includes singlet state; at $t_1$
  use $S_z$ basis for $a$, $b$ particles} \xa

\xb
\outl{Joint distribution of $S_{az}$,
  $S_{bz}$ $\ra$ $\Pr(S_{bz}\vbB S_{az})$}
\xa

\xb
\outl{Bob can calculate probabilities and conditionals, but does not know
  $S_{az}$ }
\xa

\xb
\outl{Later measurement by Bob will confirm Alice's inference}
\xa

To be more specific, use the $\FC_2$ type of framework discussed in
Sec.~\ref{sbct3.5}, with $\ket{\Psi_0}$ the initial state of Alice's apparatus
tensored with the spin singlet state of the two particles, and at the
intermediate time $t_1$ use projectors onto the orthonormal basis (with a
notation similar to \eqref{eqn2})
\begin{equation}
 \ket{z_a^+,z_b^+},\;\ket{z_a^+,z_b^-},\;\ket{z_a^-,z_b^+},\;
\ket{z_a^-,z_b^-}
\label{eqn29}
\end{equation}
of particle spin states. Their joint probability distribution at time $t_1$
given the singlet state at $t_0$ (and assuming no magnetic fields are present)
is
\begin{align}
 \Pr(S_{az}=+\hf,S_{bz}=+\hf) = 0&\quad \Pr(S_{az}=+\hf,S_{bz}=-\hf) = \hf,
\notag\\
\Pr(S_{az}=-\hf,S_{bz}=+\hf) = \hf&,\quad \Pr(S_{az}=-\hf,S_{bz}=-\hf) =0,
\label{eqn30}
\end{align}
which gives conditional probabilities
\begin{equation}
 \Pr(S_{bz}=-\hf\vbB S_{az}=+\hf) = 1 =  \Pr(S_{bz}=+\hf\vbB S_{az}=-\hf).
\label{eqn31}
\end{equation}
Since Alice knows on the basis of her measurement outcome that $S_{az}$ had the
value $+1/2$ at $t_1$, she can infer with certainty using the first equality in
\eqref{eqn31} that $S_{bz}$ had the value $-1/2$.  Note that Bob, who also
knows the protocol, can write down exactly the same probability formulas
\eqref{eqn30} and \eqref{eqn31}.  The only difference is that Alice because she
knows the outcome of her measurement can use \eqref{eqn31} to infer the value
of $S_{bz}$.  Bob of course will get this result if he carries out a later
measurement.

\xb
\outl{Wave function collapse not needed. Time of Alice's vs Bob's measurement
  unimportant}
\xa

Wave function collapse never appears in the foregoing argument.  It could be
used as a calculational tool, as in $\FC_3$ in Sec.~\ref{sbct3.5}, to compute a
conditional probability, but computational tools are not to be confused with
physical processes, and wave functions serving as pre-probabilities should be
carefully distinguished from quantum properties.  Note also that the time at
which Alice carries out a measurement, relative to when Bob does or does not
carry out a measurement, is of no importance. (See \cite{Grff11b} for a more
extended discussion of this point.) There are no mysterious influences of
Alice's measurement on Bob's particle, and thus no hint of any violation of the
principles of special relativity.

\xb \outl{What if Alice measures $S_{ax}$?  Similar, but cannot combine $S_x$
  and $S_z$ framework} \xa

And what if if Alice measures some other component of spin angular momentum,
say $S_{ax}$ rather than $S_{az}$? In that case she can use an appropriate
framework to infer the value of $S_{ax}$ before the measurement was made, and
from it deduce the value of $S_{bx}$ for particle $b$. But that framework
cannot be combined with the one which allows her to infer the earlier value of
$S_{az}$ when that is measured, since the $x$ and $z$ components of angular
momentum are incompatible quantum variables.  Alice, a competent
experimentalist, cannot measure both $S_{ax}$ and $S_{az}$ on the same particle
for there is no combined property to be measured.

\xb
\outl{Alice could have measured $S_x$ instead of $S_z$. Qm counterfactuals must
use single framework}
\xa

But even if Alice measures $S_{az}$ on this occasion, surely she could have
instead measured $S_{ax}$ on the very same particle, and in that case the
measurement would surely have revealed either $S_{ax}=+1/2$ or $-1/2$. And
therefore before the measurement the particle must have had both a definite
value of $S_{ax}$ as well as $S_{az}$. What is wrong with this argument?  The
``if\dots would'' construction betrays the presence of \emph{counterfactual}
reasoning: something actually happened, $S_{az}$ was measured, but one imagines
a different world in which $S_{ax}$ was measured instead. As discussed in
Sec.~19.4 of \cite{Grff02c}, it is important to subject counterfactual
reasoning about quantum systems to the single framework rule.  There is no
difficulty imagining a counterfactual world with distinct macroscopic
measurement setting, for these represent mutually exclusive alternatives
represented in quantum theory by commuting projectors (their product is 0).
But there is no room in the Hilbert space for simultaneous values of $S_{az}$
and $S_{ax}$, as noted earlier in Sec.~\ref{sbct3.1}.

\xb
\outl{Bell inequalities use HVs not satisfying Qm Hilbert space rules}
\xa

Here, indeed, is the point where derivations of Bell inequalities are
inconsistent with Hilbert space quantum mechanics: the inequalities are
obtained using ``hidden variables'' not subject to the rules appropriate to a
quantum Hilbert space.  The fundamental issue has nothing to do with locality,
for it already arises when considering measurements by just one party, in our
case Alice, of two incompatible physical variables belonging to
incompatible frameworks. For further discussion see \cite{Grff11b,Grff11}.

\xb
\subsection{Approximations}
\label{sbct3.8}
\xa

\xb
\outl{Is it reasonable to use approximate compatibility or consistency?}
\xa

The condition for compatibility of two frameworks for a quantum system at a
single time was stated in Sec.~\ref{sbct3.1} in terms of commutation of the
projectors from both collections.  Would not approximate commutation suffice?
The consistency condition for a family of histories at three or more times is
rather stringent; would it suffice if it were approximately satisfied?

\xb \outl{Physicists prefer simple precise theories, but often must use
  approximations} \xa

\xb
\outl{Adequacy of approximation a matter of judgment}
\xa

\xb \outl{New logic is similar: exact rules given above. Now a few words re
  approximations } \xa

As physicists we prefer to have theories stated in precise terms, especially if
the mathematical expressions are simple and ``clean'', even if in practice it
is almost always necessary to make approximations to what we believe are exact
laws in order to have a theory which can be related to the real world of
experience and experiments.  Whether a particular approximation is adequate is
an element of judgment, and it is often difficult if not impossible to provide
precise error bounds.  Quantum theory interpreted using the new logic is no
different from other physical theories in this respect, and the material given
above has deliberately been expressed in terms of exact rules.  Nonetheless,
since the new logic is intended to assist in providing a \emph{physical}
interpretation of quantum mechanics, let us add a couple of comments that may
be helpful.

\xb
\outl{Plausibility of physical equivalence of nearby Hilbert space rays}
\xa

First, it is plausible that two rays $[\psi]$ and $[\phi]$ in the Hilbert space
that are ``near'' each other in the sense that $|\inpd{\psi}{\phi}|$ is close
to 1 should have a very similar physical interpretation.  For example, the spin
of a spin-half particle has a positive component in a direction which is close
to the $z$ axis but not exactly aligned with it.  Then we can, at least for
certain purposes, think of it has having the property $S_z=+1/2$.  For example,
if the spin is measured in the $S_z$ basis the result will be  $S_z=+1/2$
most of the time, with a probability of $1-|\inpd{\psi}{\phi}|^2$ of obtaining
$S_z=-1/2$. Nor will the difference between $[\psi]$ and $[\phi]$ increase in
time under unitary time evolution.  It is considerations of this sort that
suggest that the condition that exact orthogonality of the projectors making up
a quantum sample space can be relaxed in some cases without seriously
distorting the physical interpretation. 

\xb
\outl{Plausibility of almost consistent family}
\xa

A similar intuition applies to the consistency condition \eqref{eqn14} for a
family of histories of the form \eqref{eqn7}.  If consistency is satisfied with
small errors, then it can be argued that small alterations of the projectors in
the sample space, with but small shifts in their physical interpretation, will
yield a family that exactly satisfies the consistency conditions.  In this case
the matter is not as obvious as for a simple collection of almost orthogonal
projectors, but at least it is plausible, see \cite{DwKn96}.

\xb
\section{Conceptual Difficulties of the New Logic}
\label{sct4}
\xa

\xb
\outl{CH difficulties known to RBG are listed in Table~\ref{tbl2} \& discussed
  below}
\xa

The histories approach gives rise to various conceptual difficulties, and this
section discusses those listed in Table~\ref{tbl2}.  While it may not be
complete, no important difficulty known to the author has been omitted from
this list, which helps organize the material that follows. As the new logic is
a scheme of reasoning, the issues it raises are conveniently divided into two
categories.  First, is it internally consistent, free from contradictions?
This is addressed in Sec.~\ref{sbct4.1}.  Second, assuming consistency, does it
provide a good way to think about quantum mechanics?  From the perspective of
the physicist the second question is just as important as, and perhaps even more
important than, the first: a scheme which is logically sound but does not help
us understand the world will not resolve the problems plaguing quantum
foundations. Sections~\ref{sbct4.2} to \ref{sbct4.5} address issues of the
second type, those numbered 2 to 5 in the table.

\xb
 \begin{table}[h]
 \caption{Conceptual Difficulties of the New Logic}
 \label{tbl2}
\begin{center}
 \begin{tabular}{l l l}
\hline\vspace{-1.1ex}\\
 1. &  \multicolumn{2}{l}{ Internal consistency}\\[1ex]\hline\vspace{-1.1ex}\\
 2. &    \multicolumn{2}{l}{Stochastic time development}\\
    & a. & Determinism abandoned\\
    & b. & The uniwave\\[1ex]\hline\vspace{-1.1ex}\\
 3. &    \multicolumn{2}{l}{ Framework selection}\\
    & a. & Numerous frameworks\\
    & b. & Incompatible frameworks\\
    & c. & Selection based on utility\\
    & d. & Single framework rule\\
    & e. & Choice influences reality?\\[1ex]\hline\vspace{-1.1ex}\\
 4. &    \multicolumn{2}{l}{ Particular histories}\\
    & a. & Which history occurs?\\
    & b.& Retrodiction from different measurements\\[1ex]\hline\vspace{-1.1ex}\\
 5. &    \multicolumn{2}{l}{ Truth and reality}\\
    & a. & Framework dependence of truth\\
    & b. & Unicity \\[1ex]  \hline 
\end{tabular}
\end{center}
\end{table}
\xa

\subsection{Internal consistency}
\label{sbct4.1}

\xb \outl{Rules for new logic: Choose framework. Projectors commute so ordinary
  reasoning works} \xa

Let us start by summarizing the new logic's rules for probabilistic reasoning
about the quantum world.  First, choose a quantum framework. By definition this
consists of a quantum sample space, a collection of projectors on the
appropriate quantum Hilbert space that sum to the identity, together with the
associated event algebra composed of all projectors made up of sums of sample
space projectors.  Within this framework the usual laws of \emph{classical}
probability theory and ordinary propositional logic apply without any change,
as discussed in Secs.~\ref{sbct3.1} and \ref{sbct3.2}, because the projectors
all commute with each other. And one can use the same intuition---e.g., the
sample space is a collection of mutually exclusive possibilities, one and only
of which is correct---employed in other uses of ordinary (Kolmogorov)
probability theory.  (The nontrivial issue of \emph{how} to go about choosing a
quantum framework is taken up in Sec.~\ref{sbct4.3} below.)

\xb
\outl{Single framework rule $\Ra$ consistency, by usual (Cl) arguments}
\xa

\xb
\outl{Incorrect claims that CH inconsistent ignored single framework
  rule}
\xa

The single framework rule, which is a central principle of the new logic,
prohibits combining frameworks: any sort of probabilistic argument from
premises to conclusions, including propositional logic as a special case when
probabilities are 0 or 1, must employ \emph{just one} framework. From this it
follows that arguments that prove the consistency of ordinary probabilistic or
propositional reasoning also demonstrate the consistency of quantum reasoning
based on the new logic. In particular, any contradiction that might arise
within a fixed quantum framework will also have a counterpart in standard
(classical) reasoning.  Claims made in the literature that the histories
approach is inconsistent, in the sense of leading to contradictions
\cite{Knt97,BsGh00}, are flawed in that the authors have not taken the single
framework rule seriously; see \cite{GrHr98,Grff00b}.

\xb \outl{Probabilistic arguments (Cl \& Qm) begin with a framework, sometimes
  chosen implicitly, that includes all events/histories of interest. } \xa

\xb
\outl{Some set of probabilities assumed.}
\xa

\xb
\outl{Initial data $\ra$ final conclusions:``initial'' = beginning,  ``final''
  = end of argument, not necessarily earliest and latest time}
\xa

\xb
\outl{If probs are all 0 or 1 $\lra$ ordinary logical argument}
\xa

It may help to supplement the preceding remarks with some comments about how
probabilistic reasoning is actually carried out.  Classical
applications of probability theory also begin with a framework, which is to say
a sample space and an event algebra, though this is sometimes done
implicitly---e.g., random variables are introduced without bothering to say
which space they are functions on, since the knowledgeable reader ought to know
how to construct it.  In quantum mechanics one needs to be a bit more careful,
since many---perhaps most---quantum paradoxes are constructed by combining
incompatible frameworks.  The framework chosen must, of course, include all the
events or histories one is interested in.  Next some set of probabilities are
assumed: the only strict rule for assigning them is that they must be additive
and sum to 1. Then the typical argument proceeds from some \emph{initial data},
assumed to be correct or perhaps assigned some initial probabilities, to
\emph{final conclusions}, also expressed using probabilities. If all the
probabilities are 0 or 1, one has an ordinary logical argument.  The
``initial'' in ``initial data'' refers to something assumed at the beginning of
the argument, not necessarily properties of a physical system at the earliest
time of interest, though these are often included in the initial
data. Similarly, ``final'' refers to the end of the argument, not necessarily
the latest time.

\xb \outl{Example: Family $\FC_2$ in Sec.~\ref{sbct3.5}: data =
  (initial state + measurement outcome) $\Ra$ particle state
  at intermediate time} \xa

For a specific quantum example see the discussion of measurements in
Sec.~\ref{sbct3.5} using the family $\FC_2$.  The state $[\Psi_0]$ at the
earliest time $t_0$ together with the measurement outcome at time $t_2$
constitute the initial data needed to infer a property of the particle at an
intermediate time $t_1$, which is the final conclusion.  (In this instance the
inference requires the use of probabilities obtained by applying the extended
Born rule to a closed system in a situation involving three times, so an
acceptable quantum framework must be a family of histories satisfying the
consistency conditions.)

\xb
\outl{Different frameworks with same data \& conclusions yield same
  probabilities for conclusions}
 \xa

\xb
\outl{Heuristic: use coarsest possible framework}
\xa

Often there is more than one framework which will contain the events of
specific interest along with other events, and since two such frameworks could
be incompatible, one might be concerned that they would lead to different
results, i.e., different outcome probabilities.  However, a basic consistency
rule, discussed in more detail in Ch.~16 of \cite{Grff02c}, shows that the
probabilities linking a particular set of conclusions to a specific collection
of initial data are always the same for any framework that includes both.  A
useful heuristic in constructing arguments of this sort is to employ the
coarsest possible framework, i.e., the smallest number of projectors in the
quantum sample space, that can accommodate all data and conclusions, since
adding refinements is usually more work, and there is the danger that in
constructing a complicated argument one may overlook something, such as a
consistency condition, and arrive at incorrect conclusions.

\xb
\subsection{Stochastic time development}
\label{sbct4.2}
\xa

\xb
\outl{Qm stochastic time evolution not a severe difficulty}
\xa

\xb
\outl{Can imagine a Cl nondeterministic world; Cl chaos effectively
  indeterministic }
\xa

\xb
\outl{In Qcl regime can have QM $\Ra$ determinism FAPP}
\xa

\xb
\outl{Even Einstein might have accepted Qm indeterminism}
\xa

A stochastic (probabilistic) time development in place of determinism should
not represent a serious conceptual problem. One can without difficulty imagine
a classical world in which the fundamental dynamical law has some stochastic
element, and in the regime of classical chaos even deterministic equations can
lead to behavior which is for all practical purposes indeterministic.
Furthermore, the study of quasiclassical frameworks, Sec.~\ref{sbct3.4}, shows
how fundamentally indeterministic quantum laws can, under suitable
circumstances, give rise to what is for all practical purposes deterministic
behavior for a macroscopic system.  Even Einstein might have been willing to
abandon determinism in order to achieve a theory which contains no mysterious
nonlocal influences, no longer has measurement as a fundamental principle, and
allows one to say the moon is there even when it is not being observed.

\xb
\outl{Sociological barrier: students taught to reverence Schr Eqn}
\xa

\xb
\outl{Fix this by teaching uniwave tool for finding probs, how measurements
  work }
\xa

\xb
\outl{Abandoning the uniwave $\lra$ recognizing proper role of Schr Eqn for
  finding probabilities}
\xa

There is, however, another barrier, as much sociological as scientific.
Students in their first quantum course are taught to reverence the
deterministic Schr\"odinger equation as the central principle of quantum
dynamics, whereas probabilities are treated as somewhat of an embarrassment, a
necessary evil when measurements interfere with the ``correct'' time
dependence, which however will resume again once the nasty measurement is
over. Were students at the beginning of the course taught that the unitarily
developing wave function, the uniwave, is simply a tool for calculating
probabilities, not a representation of reality, and during the course supplied
with a schematic but fully quantum description of measurements, this particular
difficulty would likely disappear.

\xb
\subsection{Framework selection}
\label{sbct4.3}
\xa

\xb
\outl{Theoretical physics uses approximate models constructed using certain
  rules}
\xa

\xb
\outl{Framework choice occurs in CP; e.g., sample space for probabilistic
  model}
\xa

\xb \outl{Will explore Qm situation by comparing with Cl relative to Liberty,
  Equality\dots} \xa

Quantum mechanics is similar to other branches of theoretical physics in that
in order to apply it to a particular system one must construct a conceptual
model following certain rules.  The choice of which model to use
depends on what one wants to discuss, but a variety of other considerations can
enter that choice.  All physical models are approximate in one way or another,
and in this sense have varying degrees of ``reality'' associated with them.
Simplifications are often introduced so as to allow a easier mathematical
analysis, or in the hopes of gaining physical insight into the problem under
discussion.  Consequently the task of choosing a framework in which to carry
out a discussion is not absent from classical physics, though it is generally
simpler than in the quantum case.  In particular, whenever probabilities are
used, there must be either an explicit or implicit choice of a sample space and
an event algebra.  Difficulties arise in the quantum case both because there
are a large number of possibilities, and also because one cannot combine
incompatible alternatives into a single description. A useful way of exploring
the quantum difficulties is to consider some classical systems, and ask which
of the principles for quantum frameworks---Liberty, Equality,
Incompatibility, Utility---introduced in Sec.~\ref{sbct3.2} have classical
analogs.

\xb
\outl{Cl analogy: Coffee cup seen from above and below}
\xa

Let us begin with an everyday classical example. It is possible to view a
coffee cup from below as well as from above, and the two perspectives give
different types of information about, or describe different aspects of, a
single object.  One can choose either, and neither perspective is more
fundamental than the other, so Liberty and Equality are represented in this
mundane example.  The Utility of each depends on what one is interested in
learning: Is there coffee in the cup?  Is there a crack on the bottom surface?
Both perspectives are \emph{compatible}: they can (in principle) be
harmoniously combined into a single, more detailed, more refined description,
containing all the information present in the views from above and below.  And
this compatibility is consistent with quantum theory: the relevant projectors,
should one be so foolish as to attempt a quantum mechanical description of the
coffee cup, will form a commuting set that is part of a quasiclassical
framework.

\xb
\outl{Compare with Sec. 3.5 example: $S_x$ prepared, $S_z$ later measured}
\xa

Contrast this with the example in Sec.~\ref{sbct3.5} where Alice prepares a
spin-half particle in a state $S_x=+1/2$ and Bob later measures it and finds
$S_z=-1/2$.  To describe the particle at an intermediate time between
preparation and measurement one can use either a consistent family of histories
which contains the value of $S_x$ at this time, or one which contains the
value of $S_z$.  Either is perfectly acceptable from the perspective of quantum
theory; there is no fundamental law that says that one should be used rather
than the other.  However, they are \emph{incompatible} with each other and
cannot  be combined.  Each family has its uses;
e.g., in addressing the question of whether the preparation was successful, or
whether the measurement apparatus performed what it was designed to do.

\xb
\outl{Compare golf ball prepared with Sx positive; Sz measured}
\xa

It is Incompatibility that most clearly marks the border between classical and
quantum physics, as can be seen by comparing the preceding example with a
situation where the spin-half particle is replaced by a golf ball which Alice
prepares with a positive $x$ component of spin angular momentum, and Bob later
measures the $z$ component and finds it is negative.  Again two valid
descriptions at the intermediate time, but now they can be combined.  Since the
typical angular momentum of a spinning golf ball is on the order of $10^{30}$
in units of $\hbar$ there is no problem constructing a quasiclassical framework
in which both $x$ and $z$ components are represented approximately with a
precision much more than adequate for all practical purposes.

\xb
\outl{Lorentz frames not a good analogy for Qm frameworks}
\xa

Another partial analogy is provided by Lorentz transformations in classical
special relativity.  The physicist is at Liberty to choose different Lorentz
frames, with none being more ``fundamental'' than another, and Utility may
determine the choice; e.g., in scattering problems there is an advantage to
using the center of mass.  However, the situation is unlike quantum mechanics
in that every Lorentz frame contains the same information as any other, since
there is a well-defined means of transforming positions and momenta between
different frames. Distinct quantum frameworks which are mutually incompatible
obviously do \emph{not} contain the same information.  (As an aside we note
that there is no particular problem in constructing a relativistic version of
the histories approach; see, e.g., \cite{Grff02b}.)

\xb
\outl{Coarse grainings in Cl Stat Mech a better analogy}
\xa

Perhaps a closer analogy is provided by classical statistical mechanics where
it is sometimes useful, for purposes of discussing irreversibility or the
origin of hydrodynamic laws, to introduce a coarse graining of the classical
phase space into nonoverlapping cells, with the coarse-grained description
providing not the actual phase point of the system, but instead the label of
the cell in which it is located.  Here the choice of coarse graining is clearly
one made by the physicist on the basis of its utility for the calculation he
has in mind, and of course no coarse graining is more ``fundamental'' than any
other.  In addition, two coarse grainings do not in general contain the same
information.  However, given any two coarse grainings there is always a common
refinement using the intersection of the cells, so with respect to
Incompatibility the analogy with (incompatible) quantum frameworks breaks down.

\xb \outl{Cl analogies imply: frameworks must be chosen, are not mutually
  exclusive, choice of framework does not influence reality} \xa

\xb
\outl{FORTRAN vs C provides partial analog of incompatibility}
\xa

To summarize the situation, classical physics provides analogies of many of the
features which need to be taken into account when thinking about quantum
frameworks.  Frameworks are not automatic: they must be chosen. There are
multiple possibilities, and alternative frameworks are not mutually exclusive
in the sense that if one is right the others must be wrong.  The physicist's
choice of framework has no influence on the reality being described, and is
generally motivated by the desire to understand or describe a particular aspect
of the system of interest.  What classical physics does not provide is a good
analogy for incompatible frameworks and the single framework rule that
prohibits combining them.  Computer languages provide a partial analogy: woe be
to the programmer who mixes FORTRAN with C.  But since a particular algorithm
can be expressed using either, this analogy, while it may be helpful, is not
exact.

\xb
\subsection{Particular histories}
\label{sbct4.4}
\xa

\xb
\outl{Objection to CH: Many families, many histories, which is the TRUE one?}
\xa

A common objection to the histories approach is that there many consistent
families of histories, and even if in each family only one elementary history
can occur, this still leaves a large number of possibilities.  How does one
know which of these histories \emph{actually} occurred?  What is the one
\emph{true} history?  What rule selects the \emph{correct} family that contains
it? 

\xb
\outl{This issue, viewed formally, is one of framework selection}
\xa

From a formal perspective the issue raised here is a particular instance of the
framework selection problem discussed above. Quantum mechanics allows many
different frameworks any one of which can be chosen by the physicist for
constructing a description, as long as they are not combined in a way that
violates the single framework rule. That prohibition includes, in the case of
consistent families of histories, the combining of incommensurate
families. Within a consistent family the elementary histories (those belonging
to the sample space) are mutually exclusive, so one and only one of them occurs
or is correct, even though quantum theory can in general only provide
probabilities for different possibilities.

\xb
\outl{Histories written by historians are not defective because they are
  different }
\xa

\xb
\outl{Should not physicist have similar liberty in choosing material?}
\xa

A formal statement is often insufficient to resolve some intuitive difficulty,
and here is where a classical analogy may be helpful.  History books written by
professional historians are often quite different, but this by itself does not
mean they are defective. The historian chooses material that
provides a coherent narrative for the time period of interest while still
remaining consistent with the facts insofar as they are known.  No one would
expect a history of the United States to cover the same territory as a history
of Great Britain.  The historian has Liberty to choose material that
best serves his purpose, and there seems no reason to deny the quantum
physicist a similar freedom.  

\xb
\outl{Histories should be consistent when discussing the same event; Qm
  histories satisfy this }
\xa

To be sure we expect different histories of the world to be consistent to the
extent that they deal with the same events, and it seems reasonable to expect
quantum descriptions to satisfy similar conditions of consistency.  And indeed
they do.  Given the same input data, without which one cannot assign
probabilities, two consistent families of histories that include this data
will always assign the same probabilities to other events that occur in both
families; this is a consequence of the internal consistency of the histories
approach discussed earlier in Sec.~\ref{sbct4.1}.

\xb \outl{Spin-half: prepare $S_x$, measure $S_z$; incompatible histories at
  intermediate time} \xa

\xb
\outl{A \& V 3-box paradox discussed in CQT. SFR removes apparent contradiction}
\xa

The example discussed earlier in which Alice prepares $S_x=+1/2$ and Bob
measures $S_z=-1/2$ may help to illustrate this point. There are two
incompatible consistent families, one containing $S_x$ and the other $S_z$ at
the intermediate time $t_1$, and in each family one draws some conclusion about
the spin angular momentum at $t_1$. The conclusions are indeed different,
but they are not contradictory, for events involving $S_x$ cannot be combined
with those involving $S_z$.  A more striking example is provided by the
three-box paradox of Aharonov and Vaidman \cite{AhVd91}, discussed in detail in
Sec.~22.5 of \cite{Grff02c}, where again the single framework rule removes the
apparent contradiction resulting from applying classical reasoning in a
situation where it violates quantum principles.

\xb
\outl{Insisting that ``there just HAS to be a SINGLE true history''. See
  following section}
\xa

To be sure, some may want to insist that ``there just \emph{has} to be a
\emph{single} history a single true story.''  This is as much a philosophical
position as a scientific objection, which does not mean it can simply be
dismissed.  The discussion in the following section is an attempt to get to the
bottom of what is here at issue.

\xb
\subsection{Truth and reality}
\label{sbct4.5}
\xa

\xb
\outl{Examples discussed earlier point to central conceptual difficulty}
\xa

\xb
\outl{Consistency of CH probabilities ensured by single framework rule}
\xa

\xb
\outl{Incompatible frameworks $\Ra$ probs relative to framework; no single prob
distribution}
\xa

The preceding examples and discussions help identify what is perhaps the
central conceptual difficulty of the new logic.  In quantum mechanics
interpreted in this way any description of nature must be formulated using a
framework of commuting Hilbert space projectors which can be assigned
probabilities in a consistent fashion.  Consistency is ensured by the single
framework rule, which prohibits combining incompatible frameworks or
incommensurate families of histories. Consequently, probability distributions
are relative to frameworks; there is not a single probability distribution that
can be used for every framework or every family of histories.

\xb
\outl{True/false $\lra$ probability 1,0; implies True/false relative to
  framework}
\xa

\xb
\outl{No single universally true state of affairs: a major block to accepting
  CH}
\xa

In a probabilistic model the limiting cases of probability 1 and 0 correspond
to statements which in propositional logic are true and false,
respectively.  This same interpretation is employed in histories quantum
mechanics.  But then ``true'' and ``false'' \emph{must be understood
relative to a framework}.  There is no single universally true state of affairs
in histories quantum mechanics.  This has undoubtedly been a major stumbling
block standing in the way of its more general acceptance by the physics
community, despite the fact that it provides a consistent resolution of all the
usual quantum paradoxes, something which cannot be said of any other
interpretation of quantum mechanics currently available.  And it seems to be
at the heart of Mermin's objection, see Sec.~\ref{sct1}.

\xb
\outl{Unicity a deeply-rooted faith called into question by QM}
\xa

Indeed, there is a deep-rooted faith or intuition, shared by scientists as well
as the ordinary man on the street, that at any point in time there is a
particular state of affairs which exists, which is ``true'', and to which every
true description of the world must conform.  No one claims to know what this
exact truth is, and indeed it must, if it exists, be beyond human knowledge.
Let us call this belief the principle of \emph{unicity}. Calling it into
question seems heretical, contrary to both common sense and sound science.
Nonetheless, in the quantum world it does not seem to be valid.  

\xb
\outl{Classical phase space provides mathematical picture of unicity}
\xa

To see how and why it breaks down, recall that a classical space can be divided
into finer and finer regions until one arrives at a single point representing
the precise state of a physical system.  All subsets of the phase space that
contain this point represent properties that are simultaneously true; their
intersection is the point itself, which is the ultimate truth.  Hence classical
mechanics provides a convenient mathematical picture for visualizing unicity,
and the enormous success of classical mechanics lends support to its validity.

\xb
\outl{Hilbert space: finest property $\lra$ 1d subspace $\lra$ point in Cl
  phase space}
\xa

\xb \outl{But then: Incompatible Qm properties completely different from CP} \xa

\xb
\outl{Unicity runs into logical problems seen by Birkhoff, vN}
\xa

\xb
\outl{CH abandons unicity; other Qm interpretations ignore the logical problem}
\xa

In the quantum Hilbert space the description of properties is ``quantized'':
subspaces have integer dimensions, and the smallest subspace representing a
property which could possibly be true has dimension 1. So this ought to be the
quantum analog of a single point in the classical phase space. But then one
finds, as discussed in detail in Sec.~\ref{sbct3.1}, that at this level the
structure of Hilbert space is significantly different from that of classical
phase space.  In particular one has properties that are incompatible---their
projectors do not commute---in a manner which is completely foreign to
classical physics.  The obvious extrapolation of classical unicity, the notion
of a single true property, runs into the logical difficulties understood by
Birkhoff and von Neumann.  Abandoning unicity, as in the new logic, may not be
the only way to solve the problem, but simply ignoring it, which is what one
finds in much modern work on quantum foundations, is not likely to result in
progress.

\xb
\outl{Qcl frameworks explain intuition behind unicity in everyday life \& why
  it fails in QM}
\xa

To put the matter in a slightly different way, if we assume that the world
is governed by classical principles we eventually run into disagreement with
experiment.  However, by assuming that it is governed by quantum principles,
with quasiclassical frameworks explaining the success of classical physics at
the macroscopic level, we can begin to understand the deep intuition that lies
behind the notion of unicity, based on everyday human experience in the
classical world.  But at the same time we can understand why and in what
circumstances this intuition breaks down.

\xb
\outl{Cl truth and reality as self evident as the fact that the earth is at 
rest}
\xa

\xb
\outl{History of science: self-evident things replaced by alternatives}
\xa

To those who claim that classical notions of truth and reality are necessary
truths, which are self evident, the appropriate response is to say that they
are just as self evident as the fact, accepted by our intellectual ancestors,
that the earth is at rest at the center of the universe.  The history of
science is marked by a set of important revolutions in thought in which things
thought to be intuitively obvious and self-evident have been replaced by
alternative explanations in better agreement with empirical observation.  Why
should quantum theory be different?

\xb
\outl{Abandoning unicity does not imply abandoning logical thought}
\xa

\xb
\outl{Need new intuition to go along with the new logic}
\xa

\xb
\outl{Abandoning unicity $\not\Ra$ giving up notion of real world ``out
  there''}
\xa

Abandoning unicity is not equivalent to abandoning logical thought, and it is
worth stressing that the histories approach to quantum interpretation, the new
quantum logic, is entirely consistent provided one pays attention to the
rules, discussed with various examples in Sec.~\ref{sct3} and summarized in
Sec.~\ref{sbct4.1} for constructing quantum descriptions.  Nor does abandoning
unicity mean that one has to give up on physical intuition about what is
``going on'' in the quantum world.  True, classical thinking is no longer
satisfactory in the quantum domain, and the physicist has to develop an
appropriate quantum intuition in its place.  This takes effort, but it is not
impossible.
Nor does abandoning unicity require giving up the notion of a real world ``out
there'', one whose existence is independent of our thoughts, wishes, and
beliefs.  What the development of quantum mechanics and its consistent
interpretation using the new logic indicates is that the certain features of
this reality differ from what was thought to be the case before quantum theory
was developed and successfully applied to understanding phenomena in the
microscopic world.

\xb
\section{Conclusion}
\label{sct5} 
\xa

\xb
\outl{Claim: new logic has successful resolved Qm conceptual 
difficulties}
\xa

\xb
\outl{It is radical break with many long accepted ideas}
\xa

\xb
\outl{It is consistent with logical thot and an independent reality}
\xa

The fundamental thesis of this paper is that the conceptual difficulties of
quantum foundations listed in Table~\ref{tbl1} and discussed in
Sec.~\ref{sct2}, can be, and in fact have been, successfully resolved using the
new quantum logic embodied in the histories approach, as summarized in
Sec.~\ref{sct3}.  The new logic, while not as radical as the older quantum
logic, still represents an important break with ideas which have long seemed
central to human thought in general and to the natural sciences in particular.
The most significant changes, and the conceptual difficulties that they in turn
give rise to, are indicated in Table~\ref{tbl2} and discussed in
Sec.~\ref{sct4}.  However novel it may seem, it is worth remembering that the
new logic is consistent both with logical thought and with an independent
reality that does not require human thought (or measurements) to bring it into
existence.

\xb
\outl{CH differs from other proposals in (i) role played by noncommutation;
(ii) stochastic time development; (iii) resolves ALL the paradoxes}
\xa

\xb
\outl{Supposed superluminal influences are simply fudge factors}
\xa

There might be yet better ways of resolving quantum conceptual difficulties
than those provided by the histories approach.  What distinguishes it from
alternative proposals at the present time is the combination of (i) the central
role played by the noncommutation of quantum operators, in particular
projectors, in the conceptual foundations of the subject; (ii) its insistence
that \emph{all} quantum time development is stochastic, not just when
measurements take place; (iii) its success in resolving not just one or two,
but \emph{all} of the standard quantum paradoxes.  In particular, the supposed
superluminal influences that have infested quantum foundations for the last 50
years and make some other interpretations of quantum mechanics difficult to
reconcile with relativity theory are absent from the histories approach; such
influences are nothing but fudge factors needed to compensate for a lack of
understanding of what quantum measurements measure, and the failure to use a
fully consistent set of quantum principles when discussing entangled states.

\xb
\outl{Accept vs reject scientific ideas a judgment call}
\xa

\xb
\outl{Cannot prove that earth moves; assuming it does makes things simpler}
\xa

\xb
\outl{Critical scrutiny of new logic is welcome. That it's radical $\not\Ra$
  it's wrong}
\xa

In the end the acceptance or rejection of a set of ideas by individual
scientists and by the scientific community is a matter of scientific judgment;
there are no overwhelming arguments that establish the proof of any scientific
theory.  That scientists today believe that the earth moves, around its axis
and around the sun, rather than lying fixed in place at the center of the
universe, is a consequence not of rigorous logical proofs of the sort that
appeal to some philosophers, but instead the fact that this way of looking at
things clears up a number of conceptual difficulties in a way much simpler and
seemingly more satisfactory than can be done by assuming the earth is fixed.
The author believes that the same is true of the new quantum logic, and
welcomes critical scrutiny by those who are willing to examine it in
detail before publishing their conclusions.  That the histories approach to
interpreting quantum theory is radical must be acknowledged.  This does not
mean it is wrong.

\section*{Acknowledgments}

The work described here is based on research supported by the National Science
Foundation through Grant PHY-1068331.

\end{document}